\documentclass[aps,pra,twocolumn,superscriptaddress]{revtex4-1}
\usepackage{mathtools}

\usepackage{booktabs}
\usepackage{multirow}

\begin{document}

\title{Direct measurement of interfacial Dzyaloshinskii-Moriya interaction in X$|$CoFeB$|$MgO heterostructures with a scanning-NV magnetometer}

\author{ I.~Gross}
\affiliation{Laboratoire Charles Coulomb, Universit\'{e} de Montpellier and CNRS, 34095 Montpellier, France}
\affiliation{Laboratoire Aim\'{e} Cotton, CNRS, Universit\'{e} Paris-Sud, ENS Cachan, Universit\'{e} Paris-Saclay, 91405 Orsay Cedex, France}
\author{L.~J.~Mart\'{\i}nez}
\affiliation{Laboratoire Charles Coulomb, Universit\'{e} de Montpellier and CNRS, 34095 Montpellier, France}
\author{ J.-P.~Tetienne}
\affiliation{Laboratoire Aim\'{e} Cotton, CNRS, Universit\'{e} Paris-Sud, ENS Cachan, Universit\'{e} Paris-Saclay, 91405 Orsay Cedex, France}
\author{T.~Hingant}
\affiliation{Laboratoire Aim\'{e} Cotton, CNRS, Universit\'{e} Paris-Sud, ENS Cachan, Universit\'{e} Paris-Saclay, 91405 Orsay Cedex, France}
\author{ J.-F.~Roch}
\affiliation{Laboratoire Aim\'{e} Cotton, CNRS, Universit\'{e} Paris-Sud, ENS Cachan, Universit\'{e} Paris-Saclay, 91405 Orsay Cedex, France}
\author{ K.~Garcia}
\affiliation{Institut d'Electronique Fondamentale, CNRS, Universit\'e Paris-Sud, Universit\'{e} Paris-Saclay, 91405 Orsay, France}
\author{R. Soucaille}
\affiliation{Institut d'Electronique Fondamentale, CNRS, Universit\'e Paris-Sud, Universit\'{e} Paris-Saclay, 91405 Orsay, France}
\author{ J. P. Adam}
\affiliation{Institut d'Electronique Fondamentale, CNRS, Universit\'e Paris-Sud, Universit\'{e} Paris-Saclay, 91405 Orsay, France}
\author{J.-V. Kim}
\affiliation{Institut d'Electronique Fondamentale, CNRS, Universit\'e Paris-Sud, Universit\'{e} Paris-Saclay, 91405 Orsay, France}
\author{S. Rohart}
\affiliation{Laboratoire de Physique des Solides, CNRS, Universit\'e Paris-Sud, Universit\'{e} Paris-Saclay, 91405 Orsay, France}
\author{A. Thiaville}
\affiliation{Laboratoire de Physique des Solides, CNRS, Universit\'e Paris-Sud, Universit\'{e} Paris-Saclay, 91405 Orsay, France}
\author{ J.~Torrejon}
\affiliation{National Institute for Materials Science, Tsukuba 305-0047, Japan}
\author{ M.~Hayashi}
\affiliation{National Institute for Materials Science, Tsukuba 305-0047, Japan}
\author{ V.~Jacques}
\email{vincent.jacques@umontpellier.fr}
\affiliation{Laboratoire Charles Coulomb, Universit\'{e} de Montpellier and CNRS, 34095 Montpellier, France}

\date{\today}

%In conclusion, we have employed scanning-NV magnetometry to locally probe the strength of the interfacial DMI in [Ta, TaN, W]$|$CoFeB$|$MgO ultrathin films. By measuring the stray field emanating from DWs in micron-long wires of such materials, we observe deviations from the Bloch profile for TaN and W underlayers that are consistent with a positive DMI value favoring right-handed chiral spin structures. While the overall trends are in accord with previous work involving current-driven wall dynamics, our study reveals important quantitative discrepancies. Moreover, our measurements suggest that the DMI constant might vary locally within a single sample. These results illustrate the importance of local probes of magnetic states and suggest certain hypotheses for extracting the DMI value from domain wall motion experiments require great care and depend strongly on assumptions made on the dynamics.

\begin{abstract}
The Dzyaloshinskii-Moriya Interaction (DMI) has recently attracted considerable interest owing to its fundamental role in the stabilization of chiral spin textures in ultrathin ferromagnets, which are interesting candidates for future spintronic technologies. Here we employ a scanning nano-magnetometer based on a single nitrogen-vacancy (NV) defect in diamond to locally probe the strength of the interfacial DMI in CoFeB$|$MgO ultrathin films grown on different heavy metal underlayers X=Ta,TaN, and W. By measuring the stray field emanating from DWs in micron-long wires of such materials, we observe deviations from the Bloch profile for TaN and W underlayers that are consistent with a positive DMI value favoring right-handed chiral spin structures. Moreover, our measurements suggest that the DMI constant might vary locally within a single sample, illustrating the importance of local probes for the study of magnetic order at the nanoscale.

\end{abstract}

\pacs{}

\maketitle

%%%%%%%%%%%%%%%%%%%%%%%%%%%%%%%%%%%%%%%%%%%%%%%%%%%%%%%

\section{Introduction}
\label{intro} 

The search for a medium that allows high information storage density combined with low power consumption, has motivated the study of low dimensional magnetic systems~\cite{Chappert2007,Parkin11042008,Wolf16112001,Fert2013}. In such materials, lowered symmetry gives rise to a new category of dominating interactions, whose interplay leads to exotic magnetization patterns~\cite{PhysRevB.78.140403,Kwon2012}. One example of such systems are magnetic thin film multilayers lacking inversion symmetry, which give rise to the Dzyaloshinskii-Moriya interaction (DMI)~\cite{dzyaloshinsky1957,PhysRevLett.4.228,MoriyaPRB}, an antisymmetric exchange interaction occurring at the interface between a ferromagnetic layer and a heavy metal substrate with large spin-orbit coupling. In ultrathin magnetic wires, interfacial DMI plays a fundamental role in the stabilization of chiral spin textures, leading to spin spirals~\cite{Bode2007,Meckler2009}, homochiral N\'eel domain walls (DWs)~\cite{HeidePRB2008,Chen2013,Thiaville2012} and magnetic skyrmions~\cite{Heinze2011,Romming2013,Jiang2015,Beach2015,Boulle2016,Moreau2016}. Since these chiral spin textures are at the heart of a number of emerging applications in spintronics~\cite{Parkin11042008,Wolf16112001,Fert2013}, it is crucial to quantify precisely the DMI strength in ultrathin ferromagnetic heterostructures. Such measurements would help to better understand the microscopic origin of interfacial DMI with the goal of controlling its strength by engineering optimized magnetic materials~\cite{HongxinPRL2015}.

A large number of experiments to date have relied on the analysis of DW motion under magnetic fields~\cite{Je2013,Hradec2014} and currents~\cite{Emori2013,Ryu2013,martinez2013,Torrejon2014} to determine the strength of the DMI in ultrathin ferromagnetic films. However, such methods rely heavily on assumptions concerning the internal spin structure of the DW and its dynamics, owing to the large number of spin torques involved in the DW dynamics~\cite{Garello2013} as well as the influence of pinning effects in the creep regime~\cite{Pizzini2015,Lavrijsen2015}. It was recently demonstrated that direct measurements of the DMI strength can be obtained by monitoring the non-reciprocal propagation of spin waves with Brillouin light spectroscopy in the Damon-Eshbach geometry~\cite{KaiPRL2015,Nembach2015,Stashkevich2015,Belmeguenai2015}. Although accurate, these measurements are always averaged over length scales of several microns, and thus can not be used to investigate local variations of the magnetic properties.

An alternative strategy consists in measuring the inner structure of DWs with the aim of observing the transition from a Bloch to a N\'eel configuration induced by the DMI~\cite{HeidePRB2008,Meckler2009,Chen2013}. In this context, it was recently shown that the nature of DWs in ultrathin ferromagnets can be inferred through quantitative stray field measurements with an atomic-sized magnetometer based on a single nitrogen-vacancy (NV) defect in diamond~\cite{Tetienne2015}. This technique, which operates under ambient conditions, enables direct measurements of the DW structure, without making any assumptions on its dynamics.

In this Letter, we use scanning NV-magnetometry to measure the strength and sign of DMI in perpendicularly magnetized X$|$CoFeB$|$MgO heterostructures. The heavy metal underlayer (X) is changed from Ta, TaN to W in order to study how the DMI strength evolves while modifying the ferromagnet/metal interface. Our results clearly indicate that the DMI is significantly enhanced when the Ta underlayer is replaced by W, leading to right-handed N\'eel walls. Our work also suggests that modifications of the underlayer thickness in the nanometer range do not translate into significant changes of the DMI strength, as expected for an interfacial effect. Finally, we reveal local modifications of the magnetic properties, which might result from inhomogeneities of the DMI strength in ultrathin ferromagnets.

 The paper is organized as follows. In Section~\ref{methods}, we first introduce the general principle of the experiment by showing how stray field measurements above a DW enable us to infer its inner structure, and thus the strength and the sign of DMI. We then describe in Section~\ref{Result} how the DMI evolves while changing the heavy metal underlayer (X=Ta,TaN,W) in X$|$CoFeB$|$MgO heterostructures. In Section~\ref{disc}, the experimental results obtained with scanning-NV magnetometry are finally compared to those obtained with other methods based either on DW motion or Brillouin light spectroscopy in the same systems. 

%%%%%%%%%%%%%%%%%%%%%%%%%%%%%%%%%%%%%%%%%%%%%%%%%%%%%%%%%

\section{Principle of the experiment}\label{methods}

We consider a DW in a thin ferromagnetic film with perpendicular magnetic anisotropy grown on top of a heavy metal substrate possessing a large spin-orbit coupling. In such a geometry with broken inversion symmetry, the strength of interfacial DMI can be large enough to modify the inner structure of the DW. The latter is characterized by the angle $\psi$ between the in-plane DW magnetization and the $x$-axis perpendicular to the DW [Fig.~1(a)]. For a Bloch DW, $\psi=\pm\pi/2$ and the magnetization rotates as a spiral while crossing the DW. A N\'eel DW rather corresponds to $\psi=0$ or $\pi$, leading to a cycloidal rotation of the magnetization. In both cases, the two possible values of $\psi$ gives the chirality (right or left) of the DW [Fig. 1(a)]. 

The interplay between the DW structure and DMI can be simply inferred by considering the surface energy density $\sigma$ [J/m$^2$] of the DW, which can be expressed as~\cite{Je2013}
\begin{equation}
\sigma= 4\sqrt{AK_{\rm eff}}+\frac{\mu_0 M_s^2 t \ln2}{\pi}\cos^2\psi-\pi D \cos\psi \ .
\label{eq:psi_DMEnergy} 
\end{equation}
Here $A$ is the exchange constant, $K_{\rm eff}$ is the effective anisotropy, $M_s$ is the saturation magnetization, $t$ is the thickness of the magnetic layer and $D$ is the micromagnetic DMI constant. The DW structure is obtained by minimizing the DW energy with respect to $\psi$, leading to
\begin{equation}
\psi=
\begin{dcases} 
\qquad 0 & \text{if} \ D> D_c \\
\pm {\rm acos}\left[\frac{D}{D_c}\right] & \text{when}  \ |D|\leq D_c\\
	\qquad \pi & \text{if} \ D< - D_c \ ,
\end{dcases}
\label{eq:psi_DMI} 
\end{equation}
where $D_c=2\mu_0M_s^2t\ln2/\pi^2 \ .$ 

\indent In the limit $|D|\ll D_c$, a Bloch DW is obtained ($\psi=\pm \pi/2$). Conversely, if $|D|\geq D_c$, the DMI strength is large enough to fully stabilize the DW into a N\'eel configuration, with a chirality fixed by the sign of $D$. In intermediate regimes, $D<D_c$, the DW moments reorient gradually towards the Bloch configuration as the DMI strength decreases.  

\begin{figure}[t]
\begin{center}
\includegraphics[width=.48\textwidth]{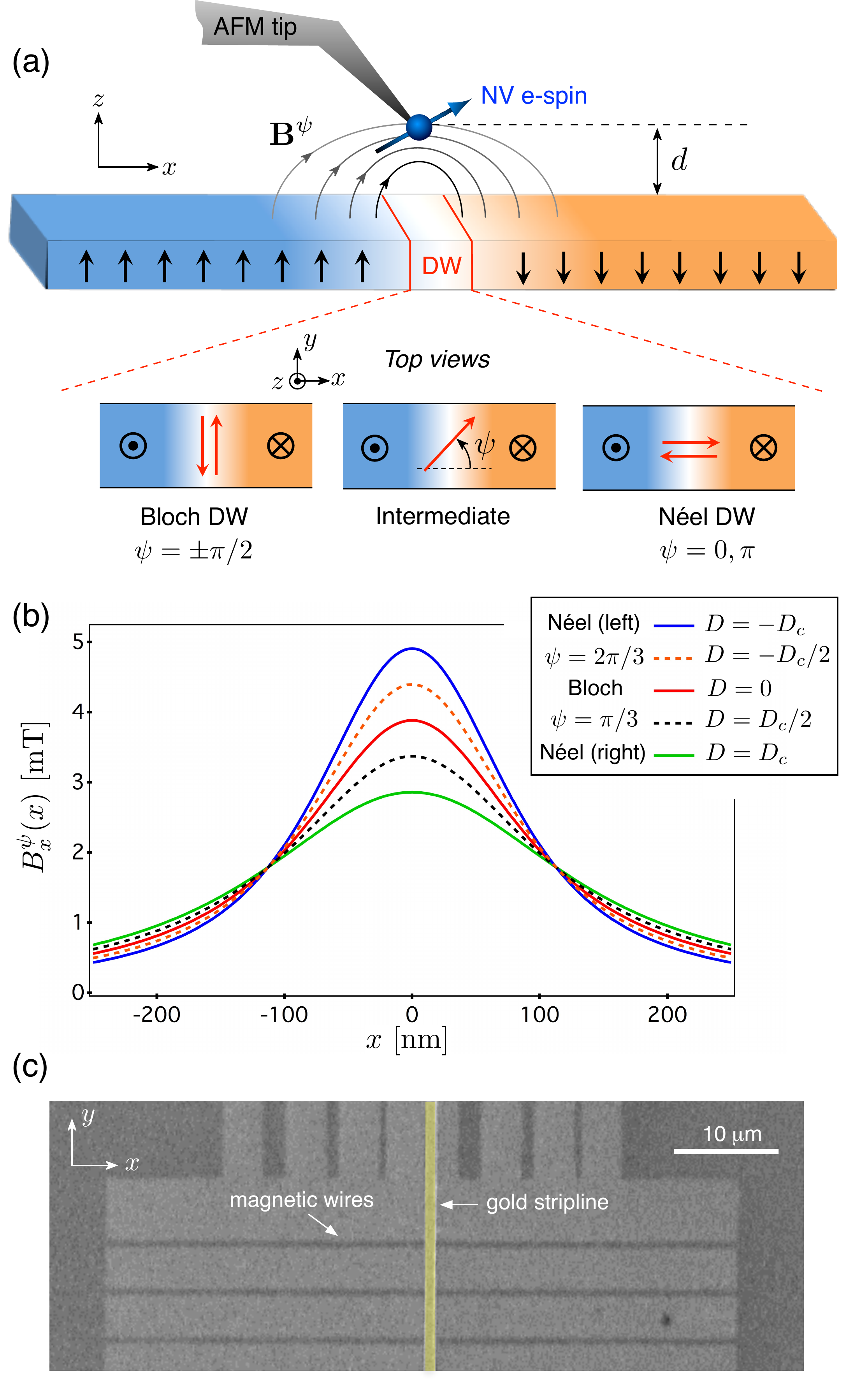}
\caption{{\bf Principle of the experiment.} (a) A scanning-NV magnetometer is used to measure the stray field $\mathbf{B}^{\psi}$ produced by a DW in a perpendicularly magnetized ferromagnetic wire (black arrows). The bottom panels show top views of the magnetization for various DW structures, characterized by the angle $\psi$ between the in-plane magnetization (red arrows) and the $x$-axis. (b) Stray field component $B_x^{\psi}(x)$ calculated for different values of the DMI strength, corresponding to different $\psi$ angles. The calculation is performed at a distance $d=100$~nm from a magnetic layer with thickness $t=1$~nm, saturation magnetization $M_s=1$~MA/m and DW width $\Delta=20$~nm. With these parameters $D_c=0.17$~mJ/m$^2$. (c) Scanning electron microscope (SEM) image of the sample used in this work, showing magnetic microwires and the gold stripline (yellow color) used both for DW nucleation and as a microwave antenna for scanning-NV magnetometry.}
\end{center}
\label{fig:fig01}
\end{figure}

It was recently shown that the DW structure, and thus the DMI strength, can be determined through quantitative measurements of the stray magnetic field above the DW~\cite{TetienneJAP2014,Tetienne2015}. Considering a one-dimensional (1D) model with an infinitely long DW along the $y$ axis [Fig. 1(a)], the stray field can be written~\cite{Tetienne2015}
\begin{equation}
\mathbf{B}^{\psi}(x)=\mathbf{B}^{\perp}(x)+\mathbf{B}^{\parallel}(x)\cos\psi \ ,
\label{eq:Bpsi} 
\end{equation}
where $\mathbf{B}^{\perp}$ (resp. $\mathbf{B}^{\parallel}$) results from the variation of the out-of-plane (resp. in plane) magnetization while crossing the DW along the $x$ direction. The stray field components at a distance $d$ above the DW located at $x=0$ are given by

\begin{equation}
\begin{dcases} 
B_x^{\bot}(x)=\frac{\mu_0 M_s t}{\pi} \frac{d}{x^2+d^2} \\
\ B_z^{\bot}(x)= -\frac{\mu_0 M_s t}{\pi} \frac{x}{x^2+d^2}
\end{dcases}
\end{equation}
and
\begin{equation}
\begin{dcases} 
B_x^{\|}(x)=\frac{1}{2}\mu_0 M_s t \Delta \frac{x^2-d^2}{(x^2+d^2)^2} \\
B_z^{\|}(x)= \mu_0 M_s t \Delta \frac{xd}{(x^2+d^2)^2} \ ,
\end{dcases}
\label{eq:stray_comp1} 
\end{equation}
where $\Delta=\sqrt{A/K_{\rm eff}}$ is the DW width parameter. We note that these simple analytic formula are valid for $d\gg(\Delta,t)$~\cite{Tetienne2015}. Figure~1(b) shows the stray field component $B^{\Psi}_{x}(x)$ calculated for different values of $D$ at a distance $d=100$~nm above the DW. This graph illustrates how local stray field measurements enable to infer the inner structure $\psi$ of the DW, from which the sign and the strength of DMI can be extracted. More precisely, a value of $D$ can be obtained as long as $|D|\leq D_c$. In ultrathin ferromagnets ($t<1$~nm), $D_c$ is typically in the range of $0.2$~mJ/m$^2$. The method is therefore sensitive to weak DMI strength. If $|D|\geq D_c$, the DW is fully stabilized in the N\'eel configuration ($\psi=0,\pi$) and stray field measurements can only give the sign of DMI and a lower bound on $D$. 

The effectiveness of this method was recently demonstrated through quantitative magnetic field imaging with a scanning-NV magnetometer~\cite{Tetienne2015}. In this experiment, a diamond nanocrystal hosting a single NV defect is grafted at the apex of an atomic force microscope (AFM) and scanned above a DW in a thin ferromagnetic wire [Fig.~1(a)]. At each point of the scan, the stray magnetic field is measured with a typical sensitivity of $10 \ \mu$T.Hz$^{-1/2}$, by recording the Zeeman shift $\Delta f_{\rm NV}$ of the NV defect electronic spin sublevels through optical detection of the magnetic resonance~\cite{Rondin2012}. In the weak magnetic field regime ($<5$mT), the Zeeman shift follows
\begin{equation}
\Delta f_{\rm NV}\approx \sqrt{(\gamma_eB_{\rm NV}/2\pi)^2 + E^{2}} \ ,
\end{equation}
where $\gamma_e/2\pi\approx 28$~GHz/T is the electron spin gyromagnetic ratio, $E$ is the transverse zero-field splitting parameter of the NV defect, which is typically in the range of few MHz, and $B_{\rm NV}$ is magnetic field projection along the NV defect quantization axis ${\bf u}_{\rm NV}$~\cite{Rondin2014}. This axis can be precisely measured independently by recording $\Delta f_{\rm NV}$ as a function of a calibrated magnetic field. Scanning-NV magnetometry then provides quantitative magnetic field distributions above DWs in thin ferromagnets, which can be directly compared with micromagnetic calculations in order to extract the inner structure of the DW.

We now briefly discuss the accuracy of this method. As illustrated by Eqs. (4) and (5), the stray field distribution above the DW strongly depends on $d$, $M_s t$ and $\Delta$. Any imprecisions on these parameters directly translates into uncertainties on the measurement of the angle $\psi$, and thus of the DMI strength. The surface density of magnetic moments $M_st$ and the distance $d$ to the magnetic layer can be measured with high accuracy by recording the magnetic field distribution across the edge of an uniformly magnetized ferromagnetic wire~\cite{Hingant2015}. Such a calibration experiment is always performed before measuring the stray field distribution above the DW. The main source of uncertainties then comes from the imperfect knowledge of the DW width, $\Delta=\sqrt{A/K_{\rm eff}}$. In this expression, although the effective anisotropy $K_{\rm eff}$ can be measured with high precision, the exchange constant $A$ has so far been difficult to determine accurately in ultra-thin films. For instance, measurements of this parameter vary from $A\approx10$~pJ/m to $A\approx30$~pJ/m in ultrathin CoFeB layers~\cite{Yamanouchi2011}. This is the main source of uncertainty in the measurement of the DW structure. More details about the uncertainty analysis can be found in Ref.~[\onlinecite{Tetienne2015}].

\section{Results}
\label{Result}

In this work, we use scanning NV-magnetometry to investigate the variations of the strength and sign of the DMI induced by modifications of the heavy metal underlayer (X) in perpendicularly magnetized X$|$CoFeB(1nm)$|$MgO heterostructures. We consider three different underlayers X=Ta, TaN and W. The films were deposited by magnetron sputtering on a Si$|$SiO$_2$(100~nm) wafer and magnetic microwires were then patterned onto the samples by using a combination of e-beam lithography and ion milling. A second step of e-beam lithography was finally performed in order to define a gold stripline, which is connected to a microwave generator and serves as an antenna to record the Zeeman shift of the NV defect magnetometer~\cite{Rondin2014}. This gold stripline was also used to nucleate DWs in the microwires through the Oersted field produced by a current pulse. The geometry of the sample is shown in Fig. 1(c) and a summary of the magnetic heterostructures studied in this work is given in Table 1. Note that the stoichiometric composition of CoFeB is different for the sample with a Ta underlayer film and that with TaN and W underlayers.

We start by examining Ta as the metal underlayer. Sample A is a Ta(5nm)$|$CoFeB(1nm)$|$MgO(2nm) trilayer stack [Fig.~2(a)]. A typical distribution of the Zeeman shift $\Delta f_{\rm NV}$ recorded with scanning-NV magnetometry above a DW in a $1.5$-$\mu$m-wide wire is shown in Fig.~2(c), together with the simultaneously recorded AFM image [Fig. 2(b)]. In order to extract the DW structure, the magnetic field distributions calculated with Eq.~(3) were first converted into Zeeman shift distributions while taking into account the NV defect quantization axis, and then compared to the experimental data~\cite{Tetienne2015}. A typical linecut of the magnetic field distribution across the DW is shown in Fig. 2(d) together with the theoretical predictions for various DW structures. Here the experimental data are very well reproduced with a purely Bloch-type DW structure. This result indicates that DMI can be safely neglected in a Ta$|$CoFeB(1nm)$|$MgO trilayer stack, as already reported in previous studies~\cite{Tetienne2015,Torrejon2014}.

\begin{figure}[t!]
\begin{center}
\includegraphics[width=0.51\textwidth]{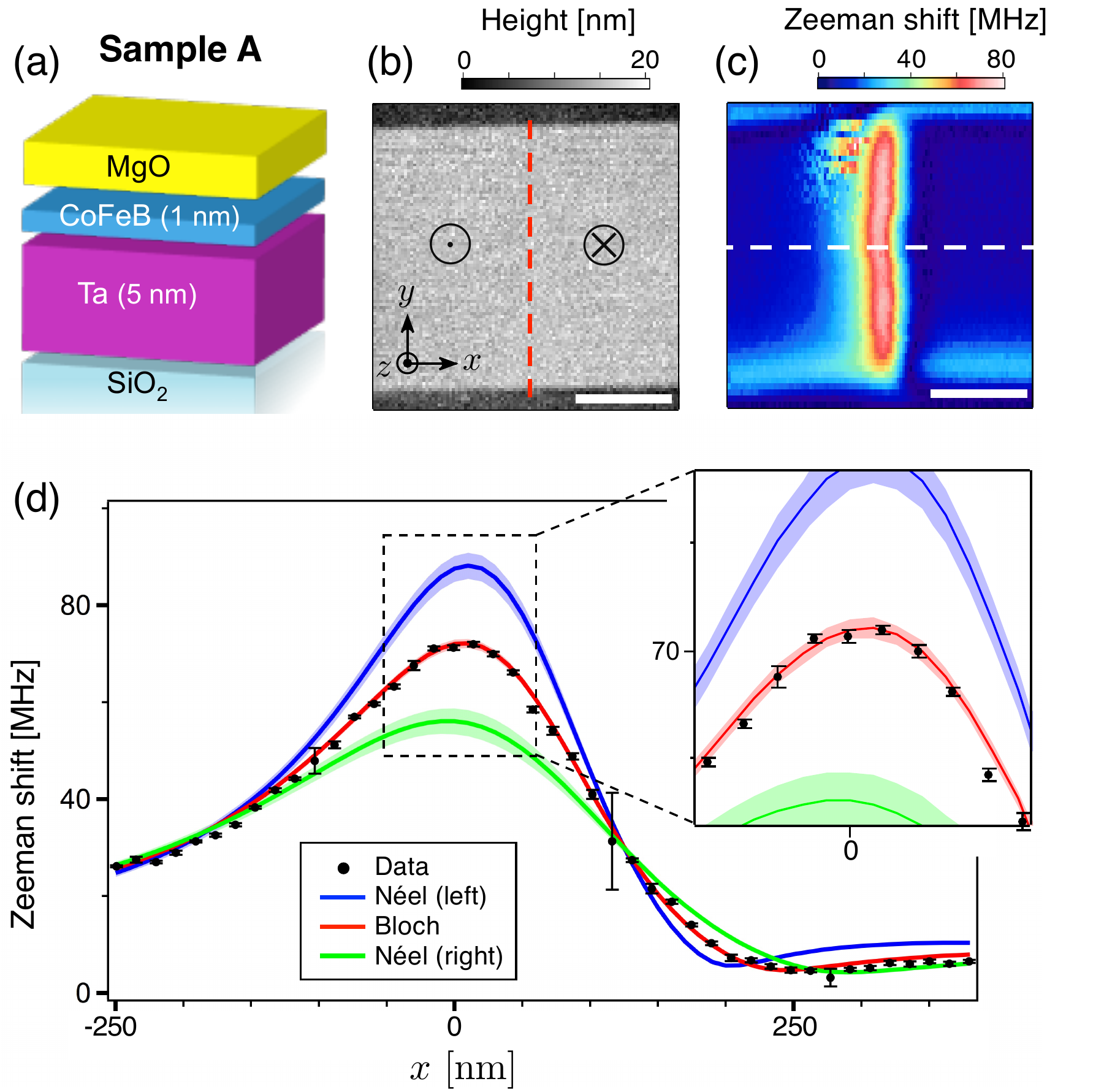}
\caption{{\bf Ta as heavy metal underlayer.} (a) Sketch of sample A. (b) AFM image and (c) corresponding Zeeman-shift distribution recorded by scanning the NV magnetometer above a DW isolated in a $1.5$-$\mu$m-wide wire of sample A. Scale bar: 500 nm. (d) Linecut extracted from the white dashed line in (c). The markers are experimental data and the solid lines are the theoretical predictions for a Bloch (red), a N\'eel left (blue) and a N\'eel right (green) DW structure. The shaded areas include all the uncertainties in the theoretical predictions, which are dominated by uncertainties on the exchange constant $A$ (see main text). The quantization axis of the NV defect ${\bf u}_{\rm NV}$ is characterized by the spherical angles $(\theta=62^{\circ},\phi=-25^{\circ})$ in the laboratory frame of reference $(x,y,z)$, the probe-to-sample distance is $d=123\pm3$~nm and $M_s t=930\pm30 \ \mu$A. We note that the Zeeman shift does not fall to zero far for the DW, because of the transverse zero-field splitting parameter of the NV defect $E$ [see Eq. (6)].}
\label{fig:fig02}
\end{center}
\end{figure}

\begin{figure*}[t]
\begin{center}
\includegraphics[width=1.01\textwidth]{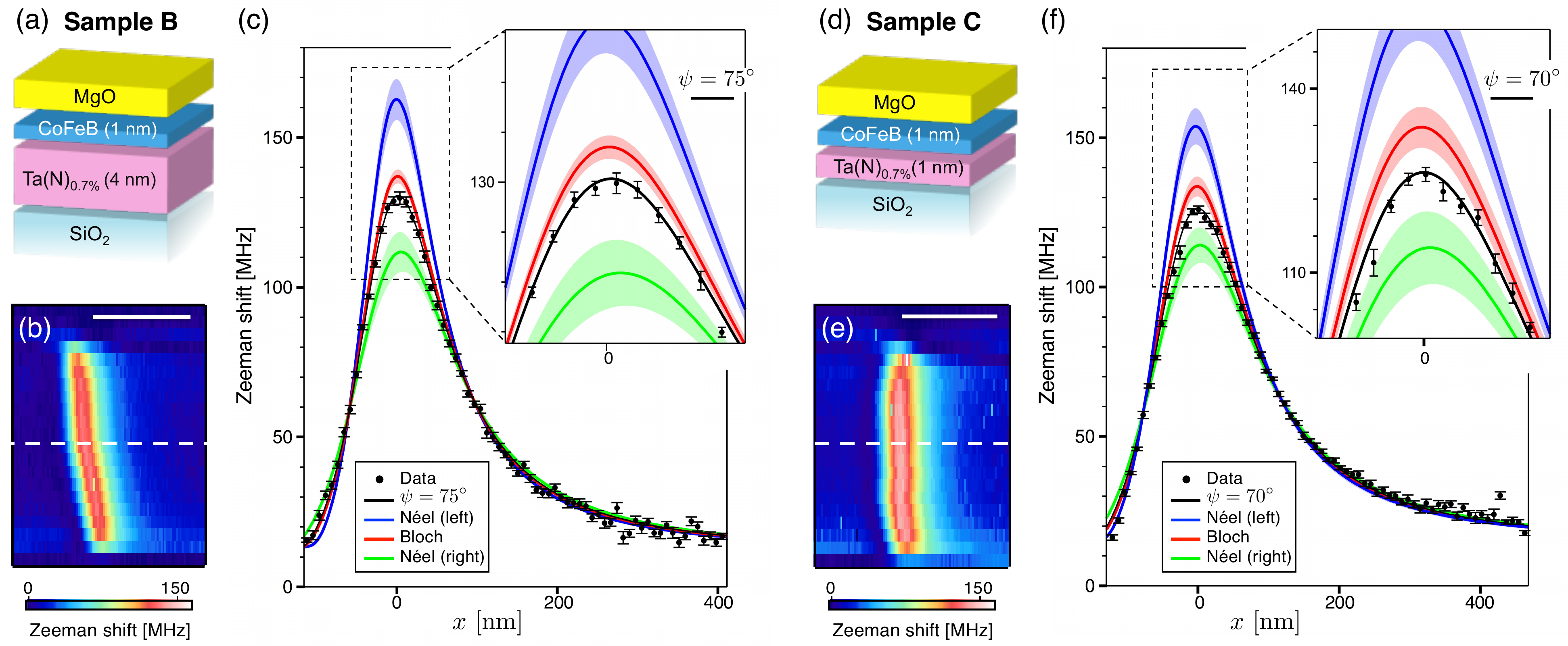}
\caption{{\bf Effect of nitrogen doping of the Ta underlayer.} (a) Sketch of Sample B. (b) Zeeman-shift distribution recorded by scanning the NV magnetometer above a DW isolated in a $1$-$\mu$m-wide wire of sample B. Scale bar: 500 nm. (c) Linecut extracted from the white dashed line in (b). The data (markers) are well reproduced by a DW structure with $\psi=75^{\circ}$ (black solid line). The quantization axis of the NV defect ${\bf u}_{\rm NV}$ is characterized by the spherical angles $(\theta=117^{\circ},\phi=12^{\circ})$ in the laboratory frame of reference $(x,y,z)$, the probe-to-sample distance is $d=61\pm6$~nm and $M_s t=800\pm80 \ \mu$A. (d) Sketch of Sample C with a 1-nm-thick TaN underlayer. (e) Zeeman-shift distribution above a DW isolated in a $1$-$\mu$m-wide wire of sample C. Scale bar: 500 nm. (f) Linecut extracted from the white dashed line in (e). The probe-to-sample distance is $d=80\pm5$~nm and $M_s t=1040\pm50 \ \mu$A. In (c) and (f), the shaded areas include all the uncertainties in the theoretical predictions.}
\label{fig:fig03}
\end{center}
\end{figure*}

It was recently shown that the magnetic properties of ultrathin CoFeB films can be significantly modified by doping the Ta underlayer with nitrogen. Such a doping leads to an enhanced interface perpendicular magnetic anisotropy~\cite{Sinha2013}. In addition, current-driven DW motion experiments have suggested that the DMI strength could also be enhanced by using a TaN underlayer~\cite{Torrejon2014}. In the following, we analyze the effect of nitrogen doping on the DMI strength by measuring the structure of DWs in Sample B, a TaN$_{0.7\%}$(4nm)$|$CoFeB(1nm)$|$MgO(2nm) trilayer stack [Fig. 3(a)]. The TaN underlayer was formed by mixing N$_2$ gas into the Ar gas atmosphere during sputtering of Ta. Here the ratio between the N$_2$ ($S_{{\rm N}_{2}}$) and the Ar ($S_{\rm Ar}$) gas flows is $Q=S_{{\rm N}_{2}}/(S_{{\rm N}_{2}}+S_{\rm Ar})=0.7 \%$, which results in an atomic composition of Ta$_{48}$N$_{52}$. The film was finally post-annealed at $300^{\circ}$C for one hour in vacuum. The magnetic field distribution recorded with scanning-NV magnetometry above a DW nucleated in a 1-$\mu$m-wide magnetic wire of sample B is shown in Fig. 3(b). Comparison with theoretical predictions indicates a small deviation of the DW structure towards a right-handed N\'eel configuration [Fig. 3(c)]. This result confirms that the DMI strength is slightly enhanced through nitrogen doping of the Ta underlayer. The experimental data are well reproduced for a DW structure with $\psi=75\pm 5^{\circ}$, corresponding to a positive DMI strength $D=0.03\pm0.01$~mJ/m$^2$. Similar results were obtained for two other DWs in Sample B. We note that this value is one order of magnitude smaller than the one inferred by Torrejon {\it et al.} in the same sample through current-driven DW motion experiments~\cite{Torrejon2014}. This discrepancy is attributed to the difficult interpretation of DW motion experiments~\cite{Je2013,Pizzini2015}, which require strong assumptions on the DW dynamics in order to quantify the current-induced spin torques at play [see Section~\ref{disc} for a detailed discussion]. Scanning-NV magnetometry rather provides a direct measurement of the DMI strength, with a DW at rest.
\begin{figure*}[t]
\begin{center}
\includegraphics[width=1.02\textwidth]{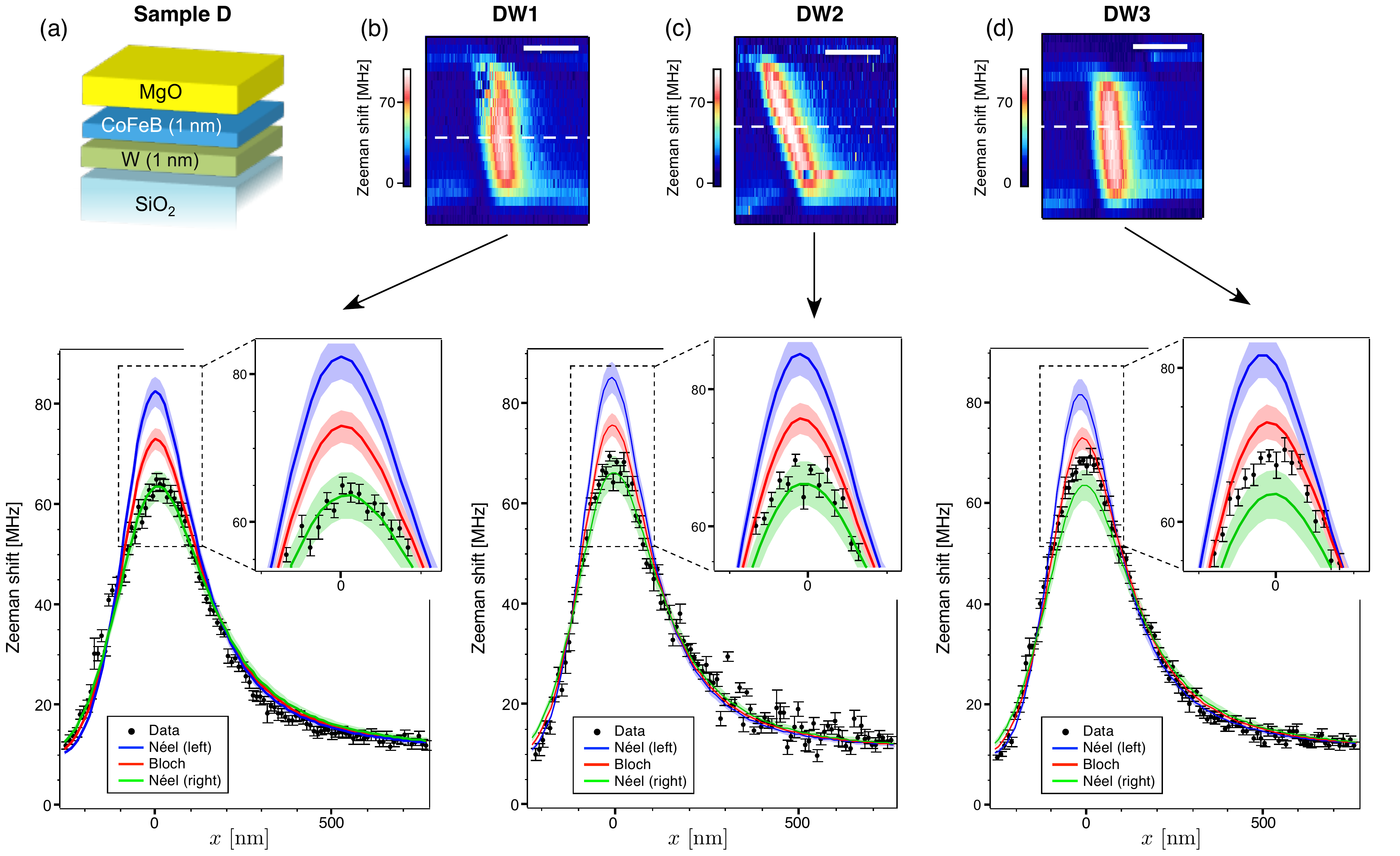}
\caption{
{\bf W as heavy metal underlayer.} (a) Sketch of Sample D. (b) to (d) Magnetic field distributions recorded above three different DWs nucleated in 1-$\mu$m-wide wire of Sample D. Scale bar: $500$~nm. The linecuts are extracted from the white dashed line in (b), (c) and (d). The shaded areas include all the uncertainties in the theoretical predictions. The quantization axis of the NV defect ${\bf u}_{\rm NV}$ is characterized by the spherical angles $(\theta=110^{\circ},\phi=12^{\circ})$ in the laboratory frame of reference $(x,y,z)$, the probe-to-sample distance is $d=114\pm10$~nm and $M_s t=820\pm100 \ \mu$A. We note that the tilt angle of the DW is taken into account in the theoretical predictions~\cite{Tetienne2015}.}
\label{fig:fig03}
\end{center}
\end{figure*}

\indent We then investigate the effect of the TaN underlayer thickness on the DMI strength. Current-induced DW motion experiments have indicated modifications of DMI with the TaN thickness~\cite{Torrejon2014}, which could appear surprising owing to the interfacial origin of DMI. In order to check these results, scanning-NV magnetometry was used to image DWs in Sample C, a TaN$_{0.7 \%}$(1~nm)$|$CoFeB(1~nm)$|$MgO(2~nm) stack [Fig. 3(d)]. Here the TaN thickness is reduced to $1$~nm. A typical magnetic field distribution recorded above a DW in a 1-$\mu$m-wide wire of Sample C is shown in Fig. 3(e). Comparison with theoretical predictions reveals once again a deviation of the DW structure towards the right-handed N\'eel configuration ($\psi=70\pm 5 ^{\circ}$), which corresponds to a DMI strength $D=0.06\pm0.02$~mJ/m$^2$ [Fig. 3(f)]. Similar results were inferred from two other DWs in Sample C. This value is close to the one measured in Sample B with a 4-nm-thick TaN underlayer. We therefore conclude that the modification of the underlayer thickness does not translate into a significant change of the DMI strength for these particular samples. This is in good agreement with the interpretation of the DMI as a surface interaction occurring at the interface between the heavy metal substrate and the magnetic layer. In contrast, we note that the value of the magnetic moment density $M_st$ changes by $\sim20\%$.

The last underlayer considered is W, in Sample D, a W(1nm)$|$CoFeB(1nm)$|$MgO(2nm) trilayer stack, which was post-annealed at $300^{\circ}$C for one hour in vacuum [Fig. 4(a)]. The magnetic field distributions recorded above three different DWs in 1-$\mu$m-wide wires of this sample are presented in Figs. 4(b)-(d). The first DW is fully stabilized in the right-handed N\'eel configuration [Fig.~4(b)]. In this case, a lower bound can be set to the DMI constant $D>D_c=0.12$~mJ/m$^2$, in good agreement with the results obtained though current-induced DW motion experiments in the same system~\cite{Torrejon2014}. The second DW also exhibits a right-handed N\'eel configuration within the uncertainty of our technique [Fig.~4(c)]. However, in the case of the third DW, the magnetic field distribution clearly indicates a DW structure lying between the Bloch and the right-handed N\'eel configurations [Fig.~4(d)]. Here, the magnetic field distribution is well reproduced for $\psi=66\pm 5^{\circ}$, corresponding to $D=0.05\pm 0.02$~mJ/m$^2$.

Such a discrepancy between experimental results obtained in different areas of Sample D can have different origins. First, it could originate from local variations of the saturation magnetization $M_s$. However, such variations would lead to localized stray field components, as reported in Ref.~\cite{Hingant2015}. Since these features were not observed in our experiment, inhomogeneities in the magnetic moment density $M_s$ can be excluded. Other possibilities are spatial variations of the effective anisotropy $K_{\rm eff}$ and/or the exchange constant $A$. Although $D_c$, hence $\psi$, would not be affected by such variations, it would however lead to a change of the DW width $\Delta$, and therefore of the stray field component $\mathbf{B}^{\parallel}$ resulting from the variation of the in-plane component of the magnetization [see Eq. (5)]. For example, the difference in stray field between a Bloch and a N\'eel-type wall would be reduced for a thinner DW. The data shown in Fig.~4(d) could then be explained if the DW width is reduced by roughly a factor of 2. This is quite unexpected owing to the good homogeneity of $M_s$ in the sample. Furthermore such a variation in $\Delta$ would lead to large variations on the DW energy, which would create strong pinning sites for the DW. This is not consistent with the low depinning fields observed in DW motion experiments in the same trilayer system~\cite{Soucaille2016}. Another possible reason explaining the discrepancy between the experimental results is a spatial variation of the DMI strength in the sample. This variation of $D$ could result from local modifications of the interface between the heavy-metal substrate and the magnetic layer, which is highly probable in a sample deposited by sputtering. These experiments illustrate how scanning-NV magnetometry enables measuring local modifications of the magnetic properties in ultrathin ferromagnets, which would be averaged out by using global techniques like Brillouin light spectroscopy~\cite{Belmeguenai2015}.

We have conclusively demonstrated that the DMI is significantly increased when the Ta underlayer is replaced by W in perpendicularly magnetized X$|$CoFeB$|$MgO heterostructures. The DMI can even be strong enough to fully stabilize the DWs onto the right-handed N\'eel configuration. We note that by changing the interface to Pt/Co, it was recently demonstrated that the sign of DMI can be reversed and the DWs then exhibit a left-handed N\'eel structure ($D<-0.1$~mJ/m$^2$)~\cite{Tetienne2015}. These experiments, which are summarized in Table 1, directly demonstrate how the strength and the sign of DMI can be tuned by engineering the interface between the heavy metal substrate and the ferromagnetic layer. 

\begin{table*}[t]
\centering
\begin{tabular}{lllccccccccc}
	\toprule[1.2pt]\\
\bf\normalsize {Name} && \bf\normalsize {Sample composition} &&\begin{tabular}[c]{@{}c@{}} $M_s$\\$[\rm A/m]\cdot 10^5$  \end{tabular} & &\begin{tabular}[c]{@{}c@{}}$\Delta$ \\ $[\rm nm]$ \end{tabular} && \begin{tabular}[c]{@{}c@{}}  {DW}\\ {structure}\\\end{tabular} && \begin{tabular}[c]{@{}c@{}} $D$\\$[\rm mJ/m^2]$\end{tabular} \\
\\
	\cmidrule[0.5pt](r){1-1}\cmidrule[0.5pt](lr){3-3} \cmidrule[0.5pt](lr){5-5}\cmidrule[0.5pt](lr){7-7}\cmidrule[0.5pt](lr){9-9}\cmidrule[0.5pt](lr){11-11}\\
\large A   && $\bf{Ta(5nm)|Co_{40}Fe_{40}B_{20}(1nm)|MgO(2nm)}$  && 9.3 $\pm$ 0.3&  & 20 $\pm$ 5 && Bloch && 0 $\pm$0.01  \\
&&&&&&& \\
\large B    && $\bf{TaN_{0.7\%}(4nm)|Co_{20}Fe_{60}B_{20}(1nm)|MgO(2nm)}$ &&  8.0$\pm$0.8 && 8 $\pm$ 2 && Bloch$^a$  &&  0.03 $\pm$ 0.01\\
&&&&&&& \\
\large C    && $\bf{TaN_{0.7\%}(1nm)|Co_{20}Fe_{60}B_{20}(1nm)|MgO(2nm)}$ && 10.4 $\pm$0.5  && 8 $\pm$ 2 && Bloch$^a$  && 0.06 $\pm$ 0.02 \\
&&&&&&& \\
\large D    && $\bf{W(1nm)|Co_{20}Fe_{60}B_{20}(1nm)|MgO(2nm)}$  && 8.2 $\pm$ 1  && 10 $\pm$ 2.5 && N\'eel-right$^b$ && $D>$0.12 \\
&&&&&&& \\
\large E \citep{Tetienne2015}  & & $\bf{Pt(3nm)|Co(0.6nm)|AlO_x(2nm)}$   && 11.2 $\pm$0.3& & 6 $\pm$1.5 && N\'eel-left && $D<$-0.1\\ 
\\
      \bottomrule[1.2pt]
\end{tabular}
\caption{{\bf Overview of the samples studied in this work.} The saturation magnetization $M_s$ is measured with scanning-NV magnetometry following the procedure described in Ref.~\cite{Hingant2015}. The DW width parameter $\Delta$ is obtained by using measurements of $K_{\rm eff}$ with vibrating sample magnetometry and assuming an exchange stiffness $A = 22\pm10$ pJ/m~\cite{Burrowes}. The strength and the sign of DMI is calculated from Eq.~(2) using the measured angle $\psi$. Sample E was measured in a previous study~\cite{Tetienne2015}. \\ $^a$ slight deviation towards a right-handed N\'eel wall.\\ $^b$ deviations within different DWs in the sample.} 
\label{tab:table}
\end{table*}

\section{Discussion}
\label{disc}
In this section, we discuss other methods that can be used to measure DMI in ultrathin ferromagnetic heterostructures, and analyze their advantages and drawbacks with respect to scanning-NV magnetometry.

The DMI can be seen as an effective in-plane chiral field $\mu_0 H_{\rm DMI}=D/M_s \Delta$ acting on the DW. A well-established way to infer $H_{\rm DMI}$, and thus the DMI strength, is based on current-driven DW motion under applied magnetic fields~\cite{Emori2013, martinez2013, Torrejon2014}. In these experiments, $H_{\rm DMI}$ is determined by measuring the longitudinal field $H_L$ at which the DW velocity goes to zero. The interpretation of the results requires to consider at least four different current-induced spin torques in the DW dynamics: the adiabatic and nonadiabatic torques related to spin-polarized current flow in the ferromagnetic layer, and the spin Hall and Rashba torques resulting from the current flow in the heavy metal substrate or at the heavy-metal/ferromagnet interface. These torques remain difficult to quantify precisely in experiments~\cite{Garello2013}. Furthermore, even if their relative strengths were known, one would also require detailed knowledge of how much of the electrical current flows in each of the metallic layers in the stack - a challenging current-in-plane problem from both experimental and theoretical points of view. Finally, the one-dimensional model used to evaluate DMI may not fully represent the current and field driven motions of DWs. In particular, the in-plane field dependence of DW velocity found experimentally does not agree well with that predicted by the model for samples with strong pinning~\cite{Torrejon2014}, which may influence the estimation of DMI. Determining the DMI field from current-driven DW motion is therefore prone to large uncertainties. %It should be noted that the current driven motion of DWs probes the DMI of a moving DW whereas the NV-microscopy reveals that of a DW at rest. 

The effective DMI field can also be inferred through field-driven DW motion. Although current-induced spin torques are not involved in this method, it still relies on the analysis of the DW dynamics. For dynamics under large fields, where the wall is driven into steady state motion, it is known that strong deformations in the lateral DW profile can appear in perpendicular anisotropy materials with weak damping. The wall structure then becomes jagged in the precessional regime where the precessing magnetization leads to dynamical Bloch and N\'eel wall states. This results in an anomalous behavior in the wall velocity versus field curve~\cite{Burrowes,Yamada}. If one were to interpret these curves with a standard one-dimensional wall model, one might mistakenly attribute the anomalous behavior to some fictitious internal field, which would be erroneous. As such, wall motion under fields alone can also be fraught with complications linked to the precise description of DW dynamics~\cite{Pizzini2015}. On the other hand in the creep regime, where the wall is pinned and motion occurs via thermal activation, the key parameter is the domain wall energy. Again, a number of strong assumptions are made to translate asymmetries in the wall velocities under applied fields to the DMI. Despite these general caveats, we remark here that recent measurements of the DMI strength obtained from domain wall creep gave similar results to those obtained in this work with scanning-NV magnetometry for the same samples~\cite{Soucaille2016}. We note that for both methods, the accuracy of the measurement is limited by the imperfect knowledge of the DW width parameter $\Delta$. 

Another strategy to determine the DMI strength relies on direct imaging of the magnetization at the wall position. This can be achieved either by spin-polarized scanning tunnelling microscopy~\cite{Meckler2009} or by spin-polarized low-energy electron microscopy~\cite{Chen2013}. However, these techniques, which require highly complex experimental apparatus with ultrahigh vaccum and a dedicated sample preparation, can hardly be used to study practical spintronic devices. Direct imaging of the DW structure can also be realized with photoemission electron microscopy combined with X-ray magnetic circular dichroism. This method was recently used to measure the chirality of magnetic skyrmions in ultrathin Pt$|$Co$|$MgO nanostructures~\cite{Boulle2016}. Finally, the wall structure can be directly inferred  through stray field imaging with a scanning-NV magnetometer, as reported in this work. This technique operates under ambient conditions and can be used to estimate the DMI strength in any type of ultrathin ferromagnetic heterostructures, without making assumptions on the DW dynamics. This is an important advantage of this method. The drawback is the limited range of DMI strength that can be measured. Indeed, as soon as $|D|\geq D_c$ the DW is fully stabilized in the N\'eel configuration and only a lower bound for $D$ can be extracted. This a common drawback of all the methods based on direct imaging of the DW structure.

All the above-mentioned techniques rely on the study of DWs nucleated in ultrathin ferromagnets. These DWs are always stabilized at pinning sites, which result from structural defects of the magnetic structure that locally lower the energy cost of a DW~\cite{Tetienne2014}. Consequently, DMI measurements based on DW properties might be systematically biased by selecting particular regions of the sample producing stable pinning sites for DWs, although we note that examination of pinned DWs has a direct technological relevance for DW-based spintronic devices. This sampling bias can be circumvented by using Brillouin light spectroscopy (BLS). Here the DMI strength is obtained by measuring frequency shifts of spin waves propagating in opposite directions of the sample~\cite{KaiPRL2015,Nembach2015,Stashkevich2015,Belmeguenai2015}. This method probes DMI over a micrometer length scale and does not rely on the presence of a DW. Local fluctuations of the magnetic properties, like the one observed in Fig.~4, are therefore averaged out. Recent BLS measurements in X$|$CoFeB$|$MgO heterostructures have systematically indicated larger $D$ values than those obtained by scanning-NV magnetometry and field-driven DW motion in the creep regime~\cite{Soucaille2016}. As already indicated above, this observation could be explained by considering that the methods relying on the study of DWs underestimate the DMI strength since the measurement is performed at stable pinning sites, corresponding to local defects of the sample which may degrade the interface. This suggests that the models used to interpret experimental results obtained with different methods still need to be refined when the aim is to obtain accurate measurements of the DMI strength. In the end, what is desirable both for physics and applications is not just the average value of $D$, but rather its complete distribution.

\section{Conclusion}\label{concl}

In conclusion, we have employed scanning-NV magnetometry to probe the strength of the interfacial DMI at the DW positions in [Ta,TaN,W]$|$CoFeB$|$MgO ultrathin films. By measuring the stray field emanating from DWs in micron-long wires of such materials, we observe deviations from the Bloch profile for TaN and W underlayers that are consistent with a positive DMI value favoring right-handed chiral spin structures. While the overall trends are in accord with previous work involving current-driven wall dynamics, our study reveals important quantitative discrepancies. Moreover, our measurements suggest that the DMI constant might vary locally within a single sample, a possibility also considered in a recent study of current-driven skyrmions motion~\cite{Beach2015}. These results illustrate the importance of local probes of magnetic states and suggest certain hypotheses for extracting the DMI value from DW motion experiments require great care and depend strongly on assumptions made on the dynamics. Given its operability under ambient conditions, we believe NV scanning magnetometry offers important new ways to study the magnetism of ultrathin films. \\

\noindent {\it Acknowledgments - }This research has been supported by the  Agence Nationale de la Recherche (France) under Contract No. ANR-14-CE26-0012 ({\sc Ultrasky}), European Union Seventh Framework Program (FP7/2007-2013) under the project {\sc Diadems} and by the European Research Council  (ERC-StG-2014, {\sc Imagine}). 

\bibliography{Igross_et_al_ref}

%merlin.mbs 2010-03-15 4.21a (PWD, AO, DPC)
\providecommand{\noopsort}[1]{}\providecommand{\singleletter}[1]{#1}%
\begin{thebibliography}{45}%
\makeatletter
\providecommand \@ifxundefined [1]{%
 \@ifx{#1\undefined}
}%
\providecommand \@ifnum [1]{%
 \ifnum #1\expandafter \@firstoftwo
 \else \expandafter \@secondoftwo
 \fi
}%
\providecommand \@ifx [1]{%
 \ifx #1\expandafter \@firstoftwo
 \else \expandafter \@secondoftwo
 \fi
}%
\providecommand \natexlab [1]{#1}%
\providecommand \enquote  [1]{``#1''}%
\providecommand \bibnamefont  [1]{#1}%
\providecommand \bibfnamefont [1]{#1}%
\providecommand \citenamefont [1]{#1}%
\providecommand \href@noop [0]{\@secondoftwo}%
\providecommand \href [0]{\begingroup \@sanitize@url \@href}%
\providecommand \@href[1]{\@@startlink{#1}\@@href}%
\providecommand \@@href[1]{\endgroup#1\@@endlink}%
\providecommand \@sanitize@url [0]{\catcode `\\12\catcode `\$12\catcode
  `\&12\catcode `\#12\catcode `\^12\catcode `\_12\catcode `\%12\relax}%
\providecommand \@@startlink[1]{}%
\providecommand \@@endlink[0]{}%
\providecommand \url  [0]{\begingroup\@sanitize@url \@url }%
\providecommand \@url [1]{\endgroup\@href {#1}{\urlprefix }}%
\providecommand \urlprefix  [0]{URL }%
\providecommand \Eprint [0]{\href }%
\@ifxundefined \urlstyle {%
  \providecommand \doi  [0]{\begingroup \@sanitize@url \@doi}%
  \providecommand \@doi [1]{\endgroup \@@startlink {\doibase
  #1}doi:\discretionary {}{}{}#1\@@endlink }%
}{%
  \providecommand \doi  [0]{doi:\discretionary{}{}{}\begingroup
  \urlstyle{rm}\Url }%
}%
\providecommand \doibase [0]{http://dx.doi.org/}%
\providecommand \Doi [0]{\begingroup \@sanitize@url \@Doi }%
\providecommand \@Doi  [1]{\endgroup\@@startlink{\doibase#1}\@@Doi}%
\providecommand \@@Doi [1]{#1\@@endlink}%
\providecommand \selectlanguage [0]{\@gobble}%
\providecommand \bibinfo  [0]{\@secondoftwo}%
\providecommand \bibfield  [0]{\@secondoftwo}%
\providecommand \translation [1]{[#1]}%
\providecommand \BibitemOpen [0]{}%
\providecommand \bibitemStop [0]{}%
\providecommand \bibitemNoStop [0]{.\EOS\space}%
\providecommand \EOS [0]{\spacefactor3000\relax}%
\providecommand \BibitemShut  [1]{\csname bibitem#1\endcsname}%
%</preamble>
\bibitem [{\citenamefont {Chappert}\ \emph {et~al.}(2007)\citenamefont
  {Chappert}, \citenamefont {Fert},\ and\ \citenamefont {Nguyen
  Van~Dau}}]{Chappert2007}%
  \BibitemOpen
  \bibfield  {author} {\bibinfo {author} {\bibfnamefont {C.}~\bibnamefont
  {Chappert}}, \bibinfo {author} {\bibfnamefont {A.}~\bibnamefont {Fert}}, \
  and\ \bibinfo {author} {\bibfnamefont {F.}~\bibnamefont {Nguyen Van~Dau}},\
  }\href@noop {} {\bibfield  {journal} {\bibinfo  {journal} {Nat. Mater.},\
  }\textbf {\bibinfo {volume} {6}},\ \bibinfo {pages} {813} (\bibinfo {year}
  {2007})}\BibitemShut {NoStop}%
\bibitem [{\citenamefont {Parkin}\ \emph {et~al.}(2008)\citenamefont {Parkin},
  \citenamefont {Hayashi},\ and\ \citenamefont {Thomas}}]{Parkin11042008}%
  \BibitemOpen
  \bibfield  {author} {\bibinfo {author} {\bibfnamefont {S.~S.~P.}\
  \bibnamefont {Parkin}}, \bibinfo {author} {\bibfnamefont {M.}~\bibnamefont
  {Hayashi}}, \ and\ \bibinfo {author} {\bibfnamefont {L.}~\bibnamefont
  {Thomas}},\ }\Doi {10.1126/science.1145799} {\bibfield  {journal} {\bibinfo
  {journal} {Science},\ }\textbf {\bibinfo {volume} {320}},\ \bibinfo {pages}
  {190} (\bibinfo {year} {2008})}\BibitemShut {NoStop}%
\bibitem [{\citenamefont {Wolf}\ \emph {et~al.}(2001)\citenamefont {Wolf},
  \citenamefont {Awschalom}, \citenamefont {Buhrman}, \citenamefont {Daughton},
  \citenamefont {von Molnar}, \citenamefont {Roukes}, \citenamefont
  {Chtchelkanova},\ and\ \citenamefont {Treger}}]{Wolf16112001}%
  \BibitemOpen
  \bibfield  {author} {\bibinfo {author} {\bibfnamefont {S.~A.}\ \bibnamefont
  {Wolf}}, \bibinfo {author} {\bibfnamefont {D.~D.}\ \bibnamefont {Awschalom}},
  \bibinfo {author} {\bibfnamefont {R.~A.}\ \bibnamefont {Buhrman}}, \bibinfo
  {author} {\bibfnamefont {J.~M.}\ \bibnamefont {Daughton}}, \bibinfo {author}
  {\bibfnamefont {S.}~\bibnamefont {von Molnar}}, \bibinfo {author}
  {\bibfnamefont {M.~L.}\ \bibnamefont {Roukes}}, \bibinfo {author}
  {\bibfnamefont {A.~Y.}\ \bibnamefont {Chtchelkanova}}, \ and\ \bibinfo
  {author} {\bibfnamefont {D.~M.}\ \bibnamefont {Treger}},\ }\Doi
  {10.1126/science.1065389} {\bibfield  {journal} {\bibinfo  {journal}
  {Science},\ }\textbf {\bibinfo {volume} {294}},\ \bibinfo {pages} {1488}
  (\bibinfo {year} {2001})}\BibitemShut {NoStop}%
\bibitem [{\citenamefont {Fert}\ \emph {et~al.}(2013)\citenamefont {Fert},
  \citenamefont {Cros},\ and\ \citenamefont {Sampaio}}]{Fert2013}%
  \BibitemOpen
  \bibfield  {author} {\bibinfo {author} {\bibfnamefont {A.}~\bibnamefont
  {Fert}}, \bibinfo {author} {\bibfnamefont {V.}~\bibnamefont {Cros}}, \ and\
  \bibinfo {author} {\bibfnamefont {J.}~\bibnamefont {Sampaio}},\ }\href
  {http://dx.doi.org/10.1038/nnano.2013.29} {\bibfield  {journal} {\bibinfo
  {journal} {Nat. Nano.},\ }\textbf {\bibinfo {volume} {8}},\ \bibinfo {pages}
  {152} (\bibinfo {year} {2013})}\BibitemShut {NoStop}%
\bibitem [{\citenamefont {Heide}\ \emph
  {et~al.}(2008){\natexlab{a}}\citenamefont {Heide}, \citenamefont
  {Bihlmayer},\ and\ \citenamefont {Bl\"ugel}}]{PhysRevB.78.140403}%
  \BibitemOpen
  \bibfield  {author} {\bibinfo {author} {\bibfnamefont {M.}~\bibnamefont
  {Heide}}, \bibinfo {author} {\bibfnamefont {G.}~\bibnamefont {Bihlmayer}}, \
  and\ \bibinfo {author} {\bibfnamefont {S.}~\bibnamefont {Bl\"ugel}},\
  }\href@noop {} {\bibfield  {journal} {\bibinfo  {journal} {Phys. Rev. B},\
  }\textbf {\bibinfo {volume} {78}},\ \bibinfo {pages} {140403} (\bibinfo
  {year} {2008}{\natexlab{a}})}\BibitemShut {NoStop}%
\bibitem [{\citenamefont {Kwon}\ \emph {et~al.}(2012)\citenamefont {Kwon},
  \citenamefont {Bu}, \citenamefont {Wu},\ and\ \citenamefont
  {Won}}]{Kwon2012}%
  \BibitemOpen
  \bibfield  {author} {\bibinfo {author} {\bibfnamefont {H.}~\bibnamefont
  {Kwon}}, \bibinfo {author} {\bibfnamefont {K.}~\bibnamefont {Bu}}, \bibinfo
  {author} {\bibfnamefont {Y.}~\bibnamefont {Wu}}, \ and\ \bibinfo {author}
  {\bibfnamefont {C.}~\bibnamefont {Won}},\ }\Doi
  {http://dx.doi.org/10.1016/j.jmmm.2012.02.044} {\bibfield  {journal}
  {\bibinfo  {journal} {J. Magn. Magn. Mater.},\ }\textbf {\bibinfo {volume}
  {324}},\ \bibinfo {pages} {2171 } (\bibinfo {year} {2012})},\ ISSN \bibinfo
  {issn} {0304-8853}\BibitemShut {NoStop}%
\bibitem [{\citenamefont {Dzyaloshinsky}(1957)}]{dzyaloshinsky1957}%
  \BibitemOpen
  \bibfield  {author} {\bibinfo {author} {\bibfnamefont {I.}~\bibnamefont
  {Dzyaloshinsky}},\ }\href@noop {} {\bibfield  {journal} {\bibinfo  {journal}
  {J. Phys. Chem. Solids},\ }\textbf {\bibinfo {volume} {4}},\ \bibinfo {pages}
  {241} (\bibinfo {year} {1957})}\BibitemShut {NoStop}%
\bibitem [{\citenamefont {Moriya}(1960){\natexlab{a}}}]{PhysRevLett.4.228}%
  \BibitemOpen
  \bibfield  {author} {\bibinfo {author} {\bibfnamefont {T.}~\bibnamefont
  {Moriya}},\ }\Doi {10.1103/PhysRevLett.4.228} {\bibfield  {journal} {\bibinfo
   {journal} {Phys. Rev. Lett.},\ }\textbf {\bibinfo {volume} {4}},\ \bibinfo
  {pages} {228} (\bibinfo {year} {1960}{\natexlab{a}})}\BibitemShut {NoStop}%
\bibitem [{\citenamefont {Moriya}(1960){\natexlab{b}}}]{MoriyaPRB}%
  \BibitemOpen
  \bibfield  {author} {\bibinfo {author} {\bibfnamefont {T.}~\bibnamefont
  {Moriya}},\ }\href@noop {} {\bibfield  {journal} {\bibinfo  {journal} {Phys.
  Rev.},\ }\textbf {\bibinfo {volume} {120}},\ \bibinfo {pages} {91} (\bibinfo
  {year} {1960}{\natexlab{b}})}\BibitemShut {NoStop}%
\bibitem [{\citenamefont {Bode}\ \emph {et~al.}(2007)\citenamefont {Bode},
  \citenamefont {Heide}, \citenamefont {von Bergmann}, \citenamefont
  {Ferriani}, \citenamefont {Heinze}, \citenamefont {Bihlmayer}, \citenamefont
  {Kubetzka}, \citenamefont {Pietzsch}, \citenamefont {Bl$\ddot{\rm u}$gel},\
  and\ \citenamefont {Wiesendanger}}]{Bode2007}%
  \BibitemOpen
  \bibfield  {author} {\bibinfo {author} {\bibfnamefont {M.}~\bibnamefont
  {Bode}}, \bibinfo {author} {\bibfnamefont {M.}~\bibnamefont {Heide}},
  \bibinfo {author} {\bibfnamefont {K.}~\bibnamefont {von Bergmann}}, \bibinfo
  {author} {\bibfnamefont {P.}~\bibnamefont {Ferriani}}, \bibinfo {author}
  {\bibfnamefont {S.}~\bibnamefont {Heinze}}, \bibinfo {author} {\bibfnamefont
  {G.}~\bibnamefont {Bihlmayer}}, \bibinfo {author} {\bibfnamefont
  {A.}~\bibnamefont {Kubetzka}}, \bibinfo {author} {\bibfnamefont
  {O.}~\bibnamefont {Pietzsch}}, \bibinfo {author} {\bibfnamefont
  {S.}~\bibnamefont {Bl$\ddot{\rm u}$gel}}, \ and\ \bibinfo {author}
  {\bibfnamefont {R.}~\bibnamefont {Wiesendanger}},\ }\href@noop {} {\bibfield
  {journal} {\bibinfo  {journal} {Nature},\ }\textbf {\bibinfo {volume}
  {447}},\ \bibinfo {pages} {190} (\bibinfo {year} {2007})}\BibitemShut
  {NoStop}%
\bibitem [{\citenamefont {Meckler}\ \emph {et~al.}(2009)\citenamefont
  {Meckler}, \citenamefont {Mikuszeit}, \citenamefont {Pre$\beta$ler},
  \citenamefont {Vedmedenko}, \citenamefont {Pietzsch},\ and\ \citenamefont
  {Wiesendanger}}]{Meckler2009}%
  \BibitemOpen
  \bibfield  {author} {\bibinfo {author} {\bibfnamefont {S.}~\bibnamefont
  {Meckler}}, \bibinfo {author} {\bibfnamefont {N.}~\bibnamefont {Mikuszeit}},
  \bibinfo {author} {\bibfnamefont {A.}~\bibnamefont {Pre$\beta$ler}}, \bibinfo
  {author} {\bibfnamefont {E.~Y.}\ \bibnamefont {Vedmedenko}}, \bibinfo
  {author} {\bibfnamefont {O.}~\bibnamefont {Pietzsch}}, \ and\ \bibinfo
  {author} {\bibfnamefont {R.}~\bibnamefont {Wiesendanger}},\ }\href@noop {}
  {\bibfield  {journal} {\bibinfo  {journal} {Phys. Rev. Lett.},\ }\textbf
  {\bibinfo {volume} {103}},\ \bibinfo {pages} {157201} (\bibinfo {year}
  {2009})}\BibitemShut {NoStop}%
\bibitem [{\citenamefont {Heide}\ \emph
  {et~al.}(2008){\natexlab{b}}\citenamefont {Heide}, \citenamefont
  {Bihlmayer},\ and\ \citenamefont {Bl$\ddot{\rm u}$gel}}]{HeidePRB2008}%
  \BibitemOpen
  \bibfield  {author} {\bibinfo {author} {\bibfnamefont {M.}~\bibnamefont
  {Heide}}, \bibinfo {author} {\bibfnamefont {G.}~\bibnamefont {Bihlmayer}}, \
  and\ \bibinfo {author} {\bibfnamefont {S.}~\bibnamefont {Bl$\ddot{\rm
  u}$gel}},\ }\href@noop {} {\bibfield  {journal} {\bibinfo  {journal} {Phys.
  Rev. B},\ }\textbf {\bibinfo {volume} {78}},\ \bibinfo {pages} {140403(R)}
  (\bibinfo {year} {2008}{\natexlab{b}})}\BibitemShut {NoStop}%
\bibitem [{\citenamefont {Chen}\ \emph {et~al.}(2013)\citenamefont {Chen},
  \citenamefont {Ma}, \citenamefont {N'Diaye}, \citenamefont {Kwon},
  \citenamefont {Won}, \citenamefont {Wu},\ and\ \citenamefont
  {Schmid}}]{Chen2013}%
  \BibitemOpen
  \bibfield  {author} {\bibinfo {author} {\bibfnamefont {G.}~\bibnamefont
  {Chen}}, \bibinfo {author} {\bibfnamefont {T.}~\bibnamefont {Ma}}, \bibinfo
  {author} {\bibfnamefont {A.~T.}\ \bibnamefont {N'Diaye}}, \bibinfo {author}
  {\bibfnamefont {H.}~\bibnamefont {Kwon}}, \bibinfo {author} {\bibfnamefont
  {C.}~\bibnamefont {Won}}, \bibinfo {author} {\bibfnamefont {Y.}~\bibnamefont
  {Wu}}, \ and\ \bibinfo {author} {\bibfnamefont {A.~K.}\ \bibnamefont
  {Schmid}},\ }\href@noop {} {\bibfield  {journal} {\bibinfo  {journal} {Nat.
  Commun.},\ }\textbf {\bibinfo {volume} {4}} (\bibinfo {year}
  {2013})}\BibitemShut {NoStop}%
\bibitem [{\citenamefont {Thiaville}\ \emph {et~al.}(2012)\citenamefont
  {Thiaville}, \citenamefont {Rohart}, \citenamefont {Ju\'{e}}, \citenamefont
  {Cros},\ and\ \citenamefont {Fert}}]{Thiaville2012}%
  \BibitemOpen
  \bibfield  {author} {\bibinfo {author} {\bibfnamefont {A.}~\bibnamefont
  {Thiaville}}, \bibinfo {author} {\bibfnamefont {S.}~\bibnamefont {Rohart}},
  \bibinfo {author} {\bibfnamefont {E.}~\bibnamefont {Ju\'{e}}}, \bibinfo
  {author} {\bibfnamefont {V.}~\bibnamefont {Cros}}, \ and\ \bibinfo {author}
  {\bibfnamefont {A.}~\bibnamefont {Fert}},\ }\href@noop {} {\bibfield
  {journal} {\bibinfo  {journal} {Europhys. Lett.},\ }\textbf {\bibinfo
  {volume} {100}},\ \bibinfo {pages} {57002} (\bibinfo {year}
  {2012})}\BibitemShut {NoStop}%
\bibitem [{\citenamefont {Heinze}\ \emph {et~al.}(2011)\citenamefont {Heinze},
  \citenamefont {von Bergmann}, \citenamefont {Menzel}, \citenamefont {Brede},
  \citenamefont {Kubetzka}, \citenamefont {Wiesendanger}, \citenamefont
  {Bihlmayer},\ and\ \citenamefont {Bl$\ddot{\rm u}$gel}}]{Heinze2011}%
  \BibitemOpen
  \bibfield  {author} {\bibinfo {author} {\bibfnamefont {S.}~\bibnamefont
  {Heinze}}, \bibinfo {author} {\bibfnamefont {K.}~\bibnamefont {von
  Bergmann}}, \bibinfo {author} {\bibfnamefont {M.}~\bibnamefont {Menzel}},
  \bibinfo {author} {\bibfnamefont {J.}~\bibnamefont {Brede}}, \bibinfo
  {author} {\bibfnamefont {A.}~\bibnamefont {Kubetzka}}, \bibinfo {author}
  {\bibfnamefont {R.}~\bibnamefont {Wiesendanger}}, \bibinfo {author}
  {\bibfnamefont {G.}~\bibnamefont {Bihlmayer}}, \ and\ \bibinfo {author}
  {\bibfnamefont {S.}~\bibnamefont {Bl$\ddot{\rm u}$gel}},\ }\href@noop {}
  {\bibfield  {journal} {\bibinfo  {journal} {Nat. Phys.},\ }\textbf {\bibinfo
  {volume} {7}},\ \bibinfo {pages} {713} (\bibinfo {year} {2011})}\BibitemShut
  {NoStop}%
\bibitem [{\citenamefont {Romming}\ \emph {et~al.}(2013)\citenamefont
  {Romming}, \citenamefont {Hanneken}, \citenamefont {Menzel}, \citenamefont
  {Bickel}, \citenamefont {Wolter}, \citenamefont {von Bergmann}, \citenamefont
  {Kubetzka},\ and\ \citenamefont {Wiesendanger}}]{Romming2013}%
  \BibitemOpen
  \bibfield  {author} {\bibinfo {author} {\bibfnamefont {N.}~\bibnamefont
  {Romming}}, \bibinfo {author} {\bibfnamefont {C.}~\bibnamefont {Hanneken}},
  \bibinfo {author} {\bibfnamefont {M.}~\bibnamefont {Menzel}}, \bibinfo
  {author} {\bibfnamefont {J.~E.}\ \bibnamefont {Bickel}}, \bibinfo {author}
  {\bibfnamefont {B.}~\bibnamefont {Wolter}}, \bibinfo {author} {\bibfnamefont
  {K.}~\bibnamefont {von Bergmann}}, \bibinfo {author} {\bibfnamefont
  {A.}~\bibnamefont {Kubetzka}}, \ and\ \bibinfo {author} {\bibfnamefont
  {R.}~\bibnamefont {Wiesendanger}},\ }\href@noop {} {\bibfield  {journal}
  {\bibinfo  {journal} {Science},\ }\textbf {\bibinfo {volume} {341}},\
  \bibinfo {pages} {6146} (\bibinfo {year} {2013})}\BibitemShut {NoStop}%
\bibitem [{\citenamefont {Jiang}\ \emph {et~al.}(2015)\citenamefont {Jiang},
  \citenamefont {Upadhyaya}, \citenamefont {Zhang}, \citenamefont {Yu},
  \citenamefont {Jungfleisch}, \citenamefont {Fradin}, \citenamefont {Pearson},
  \citenamefont {Tserkovnyak}, \citenamefont {Wang}, \citenamefont {Heinonen},
  \citenamefont {te~Velthuis},\ and\ \citenamefont {Hoffmann}}]{Jiang2015}%
  \BibitemOpen
  \bibfield  {author} {\bibinfo {author} {\bibfnamefont {W.}~\bibnamefont
  {Jiang}}, \bibinfo {author} {\bibfnamefont {P.}~\bibnamefont {Upadhyaya}},
  \bibinfo {author} {\bibfnamefont {W.}~\bibnamefont {Zhang}}, \bibinfo
  {author} {\bibfnamefont {G.}~\bibnamefont {Yu}}, \bibinfo {author}
  {\bibfnamefont {M.~B.}\ \bibnamefont {Jungfleisch}}, \bibinfo {author}
  {\bibfnamefont {F.~Y.}\ \bibnamefont {Fradin}}, \bibinfo {author}
  {\bibfnamefont {J.~E.}\ \bibnamefont {Pearson}}, \bibinfo {author}
  {\bibfnamefont {Y.}~\bibnamefont {Tserkovnyak}}, \bibinfo {author}
  {\bibfnamefont {K.~L.}\ \bibnamefont {Wang}}, \bibinfo {author}
  {\bibfnamefont {O.}~\bibnamefont {Heinonen}}, \bibinfo {author}
  {\bibfnamefont {S.~G.~E.}\ \bibnamefont {te~Velthuis}}, \ and\ \bibinfo
  {author} {\bibfnamefont {A.}~\bibnamefont {Hoffmann}},\ }\Doi
  {10.1126/science.aaa1442} {\bibfield  {journal} {\bibinfo  {journal}
  {Science},\ }\textbf {\bibinfo {volume} {349}},\ \bibinfo {pages} {283}
  (\bibinfo {year} {2015})}\BibitemShut {NoStop}%
\bibitem [{\citenamefont {Woo}\ \emph {et~al.}(2016)\citenamefont {Woo},
  \citenamefont {Litzius}, \citenamefont {Kr$\ddot{\rm u}$ger}, \citenamefont
  {Im}, \citenamefont {Caretta}, \citenamefont {Richter}, \citenamefont {Mann},
  \citenamefont {Krone}, \citenamefont {Reeve}, \citenamefont {Weigand},
  \citenamefont {Agrawal}, \citenamefont {Fischer}, \citenamefont {Kl$\ddot{\rm
  a}$ui},\ and\ \citenamefont {Beach}}]{Beach2015}%
  \BibitemOpen
  \bibfield  {author} {\bibinfo {author} {\bibfnamefont {S.}~\bibnamefont
  {Woo}}, \bibinfo {author} {\bibfnamefont {K.}~\bibnamefont {Litzius}},
  \bibinfo {author} {\bibfnamefont {B.}~\bibnamefont {Kr$\ddot{\rm u}$ger}},
  \bibinfo {author} {\bibfnamefont {M.-Y.}\ \bibnamefont {Im}}, \bibinfo
  {author} {\bibfnamefont {L.}~\bibnamefont {Caretta}}, \bibinfo {author}
  {\bibfnamefont {K.}~\bibnamefont {Richter}}, \bibinfo {author} {\bibfnamefont
  {M.}~\bibnamefont {Mann}}, \bibinfo {author} {\bibfnamefont {A.}~\bibnamefont
  {Krone}}, \bibinfo {author} {\bibfnamefont {R.}~\bibnamefont {Reeve}},
  \bibinfo {author} {\bibfnamefont {M.}~\bibnamefont {Weigand}}, \bibinfo
  {author} {\bibfnamefont {P.}~\bibnamefont {Agrawal}}, \bibinfo {author}
  {\bibfnamefont {P.}~\bibnamefont {Fischer}}, \bibinfo {author} {\bibfnamefont
  {M.}~\bibnamefont {Kl$\ddot{\rm a}$ui}}, \ and\ \bibinfo {author}
  {\bibfnamefont {G.~S.~D.}\ \bibnamefont {Beach}},\ }\href@noop {} {\bibfield
  {journal} {\bibinfo  {journal} {Nat. Mater.},\ }\textbf {\bibinfo {volume}
  {15}},\ \bibinfo {pages} {501} (\bibinfo {year} {2016})}\BibitemShut
  {NoStop}%
\bibitem [{\citenamefont {Boulle}\ \emph {et~al.}(2016)\citenamefont {Boulle},
  \citenamefont {Vogel}, \citenamefont {Yang}, \citenamefont {Pizzini},
  \citenamefont {de~Souza~Chaves}, \citenamefont {Locatelli}, \citenamefont
  {Mentes}, \citenamefont {Sala}, \citenamefont {Buda-Prejbeanu}, \citenamefont
  {Klein}, \citenamefont {Belmeguenai}, \citenamefont {Roussign\'e},
  \citenamefont {Stashkevich}, \citenamefont {Ch\'erif}, \citenamefont
  {Aballe}, \citenamefont {Foerster}, \citenamefont {Chshiev}, \citenamefont
  {Auffret}, \citenamefont {Miron},\ and\ \citenamefont {Gaudin}}]{Boulle2016}%
  \BibitemOpen
  \bibfield  {author} {\bibinfo {author} {\bibfnamefont {O.}~\bibnamefont
  {Boulle}}, \bibinfo {author} {\bibfnamefont {J.}~\bibnamefont {Vogel}},
  \bibinfo {author} {\bibfnamefont {H.}~\bibnamefont {Yang}}, \bibinfo {author}
  {\bibfnamefont {S.}~\bibnamefont {Pizzini}}, \bibinfo {author} {\bibfnamefont
  {D.}~\bibnamefont {de~Souza~Chaves}}, \bibinfo {author} {\bibfnamefont
  {A.}~\bibnamefont {Locatelli}}, \bibinfo {author} {\bibfnamefont {T.~O.}\
  \bibnamefont {Mentes}}, \bibinfo {author} {\bibfnamefont {A.}~\bibnamefont
  {Sala}}, \bibinfo {author} {\bibfnamefont {L.~D.}\ \bibnamefont
  {Buda-Prejbeanu}}, \bibinfo {author} {\bibfnamefont {O.}~\bibnamefont
  {Klein}}, \bibinfo {author} {\bibfnamefont {M.}~\bibnamefont {Belmeguenai}},
  \bibinfo {author} {\bibfnamefont {Y.}~\bibnamefont {Roussign\'e}}, \bibinfo
  {author} {\bibfnamefont {A.}~\bibnamefont {Stashkevich}}, \bibinfo {author}
  {\bibfnamefont {S.~M.}\ \bibnamefont {Ch\'erif}}, \bibinfo {author}
  {\bibfnamefont {L.}~\bibnamefont {Aballe}}, \bibinfo {author} {\bibfnamefont
  {M.}~\bibnamefont {Foerster}}, \bibinfo {author} {\bibfnamefont
  {M.}~\bibnamefont {Chshiev}}, \bibinfo {author} {\bibfnamefont
  {S.}~\bibnamefont {Auffret}}, \bibinfo {author} {\bibfnamefont {I.~M.}\
  \bibnamefont {Miron}}, \ and\ \bibinfo {author} {\bibfnamefont
  {G.}~\bibnamefont {Gaudin}},\ }\href@noop {} {\bibfield  {journal} {\bibinfo
  {journal} {Nat. Nano.},\ }\textbf {\bibinfo {volume} {11}},\ \bibinfo {pages}
  {449} (\bibinfo {year} {2016})}\BibitemShut {NoStop}%
\bibitem [{\citenamefont {Moreau-Luchaire}\ \emph {et~al.}(2016)\citenamefont
  {Moreau-Luchaire}, \citenamefont {Moutafis}, \citenamefont {Reyren},
  \citenamefont {Sampaio}, \citenamefont {Vaz}, \citenamefont {Horne},
  \citenamefont {Bouzehouane}, \citenamefont {Garcia}, \citenamefont
  {Deranlot}, \citenamefont {Warnicke}, \citenamefont {Wohlh$\ddot{\rm u}$ter},
  \citenamefont {George}, \citenamefont {Weigand}, \citenamefont {Raabe},
  \citenamefont {Cros},\ and\ \citenamefont {Fert}}]{Moreau2016}%
  \BibitemOpen
  \bibfield  {author} {\bibinfo {author} {\bibfnamefont {C.}~\bibnamefont
  {Moreau-Luchaire}}, \bibinfo {author} {\bibfnamefont {C.}~\bibnamefont
  {Moutafis}}, \bibinfo {author} {\bibfnamefont {N.}~\bibnamefont {Reyren}},
  \bibinfo {author} {\bibfnamefont {J.}~\bibnamefont {Sampaio}}, \bibinfo
  {author} {\bibfnamefont {C.~A.~F.}\ \bibnamefont {Vaz}}, \bibinfo {author}
  {\bibfnamefont {N.~V.}\ \bibnamefont {Horne}}, \bibinfo {author}
  {\bibfnamefont {K.}~\bibnamefont {Bouzehouane}}, \bibinfo {author}
  {\bibfnamefont {K.}~\bibnamefont {Garcia}}, \bibinfo {author} {\bibfnamefont
  {C.}~\bibnamefont {Deranlot}}, \bibinfo {author} {\bibfnamefont
  {P.}~\bibnamefont {Warnicke}}, \bibinfo {author} {\bibfnamefont
  {P.}~\bibnamefont {Wohlh$\ddot{\rm u}$ter}}, \bibinfo {author} {\bibfnamefont
  {J.-M.}\ \bibnamefont {George}}, \bibinfo {author} {\bibfnamefont
  {M.}~\bibnamefont {Weigand}}, \bibinfo {author} {\bibfnamefont
  {J.}~\bibnamefont {Raabe}}, \bibinfo {author} {\bibfnamefont
  {V.}~\bibnamefont {Cros}}, \ and\ \bibinfo {author} {\bibfnamefont
  {A.}~\bibnamefont {Fert}},\ }\href@noop {} {\bibfield  {journal} {\bibinfo
  {journal} {Nat. Nano.},\ }\textbf {\bibinfo {volume} {11}},\ \bibinfo {pages}
  {444} (\bibinfo {year} {2016})}\BibitemShut {NoStop}%
\bibitem [{\citenamefont {Yang}\ \emph {et~al.}(2015)\citenamefont {Yang},
  \citenamefont {Thiaville}, \citenamefont {Rohart}, \citenamefont {Fert},\
  and\ \citenamefont {Chshiev}}]{HongxinPRL2015}%
  \BibitemOpen
  \bibfield  {author} {\bibinfo {author} {\bibfnamefont {H.}~\bibnamefont
  {Yang}}, \bibinfo {author} {\bibfnamefont {A.}~\bibnamefont {Thiaville}},
  \bibinfo {author} {\bibfnamefont {S.}~\bibnamefont {Rohart}}, \bibinfo
  {author} {\bibfnamefont {A.}~\bibnamefont {Fert}}, \ and\ \bibinfo {author}
  {\bibfnamefont {M.}~\bibnamefont {Chshiev}},\ }\href@noop {} {\bibfield
  {journal} {\bibinfo  {journal} {Phys. Rev. Lett.},\ }\textbf {\bibinfo
  {volume} {115}},\ \bibinfo {pages} {267210} (\bibinfo {year}
  {2015})}\BibitemShut {NoStop}%
\bibitem [{\citenamefont {Je}\ \emph {et~al.}(2013)\citenamefont {Je},
  \citenamefont {Kim}, \citenamefont {Yoo}, \citenamefont {Min}, \citenamefont
  {Lee},\ and\ \citenamefont {Choe}}]{Je2013}%
  \BibitemOpen
  \bibfield  {author} {\bibinfo {author} {\bibfnamefont {S.-G.}\ \bibnamefont
  {Je}}, \bibinfo {author} {\bibfnamefont {D.-H.}\ \bibnamefont {Kim}},
  \bibinfo {author} {\bibfnamefont {S.-C.}\ \bibnamefont {Yoo}}, \bibinfo
  {author} {\bibfnamefont {B.-C.}\ \bibnamefont {Min}}, \bibinfo {author}
  {\bibfnamefont {K.-J.}\ \bibnamefont {Lee}}, \ and\ \bibinfo {author}
  {\bibfnamefont {S.-B.}\ \bibnamefont {Choe}},\ }\Doi
  {10.1103/PhysRevB.88.214401} {\bibfield  {journal} {\bibinfo  {journal}
  {Phys. Rev. B},\ }\textbf {\bibinfo {volume} {88}},\ \bibinfo {pages}
  {214401} (\bibinfo {year} {2013})}\BibitemShut {NoStop}%
\bibitem [{\citenamefont {Hrabec}\ \emph {et~al.}(2014)\citenamefont {Hrabec},
  \citenamefont {Porter}, \citenamefont {Wells}, \citenamefont {Benitez},
  \citenamefont {Burnell}, \citenamefont {McVitie}, \citenamefont {McGrouther},
  \citenamefont {Moore},\ and\ \citenamefont {Marrows}}]{Hradec2014}%
  \BibitemOpen
  \bibfield  {author} {\bibinfo {author} {\bibfnamefont {A.}~\bibnamefont
  {Hrabec}}, \bibinfo {author} {\bibfnamefont {N.~A.}\ \bibnamefont {Porter}},
  \bibinfo {author} {\bibfnamefont {A.}~\bibnamefont {Wells}}, \bibinfo
  {author} {\bibfnamefont {M.~J.}\ \bibnamefont {Benitez}}, \bibinfo {author}
  {\bibfnamefont {G.}~\bibnamefont {Burnell}}, \bibinfo {author} {\bibfnamefont
  {S.}~\bibnamefont {McVitie}}, \bibinfo {author} {\bibfnamefont
  {D.}~\bibnamefont {McGrouther}}, \bibinfo {author} {\bibfnamefont {T.~A.}\
  \bibnamefont {Moore}}, \ and\ \bibinfo {author} {\bibfnamefont {C.~H.}\
  \bibnamefont {Marrows}},\ }\href@noop {} {\bibfield  {journal} {\bibinfo
  {journal} {Phys. Rev. B},\ }\textbf {\bibinfo {volume} {90}},\ \bibinfo
  {pages} {020402} (\bibinfo {year} {2014})}\BibitemShut {NoStop}%
\bibitem [{\citenamefont {Emori}\ \emph {et~al.}(2013)\citenamefont {Emori},
  \citenamefont {Bauer}, \citenamefont {Ahn}, \citenamefont {Martinez},\ and\
  \citenamefont {Beach}}]{Emori2013}%
  \BibitemOpen
  \bibfield  {author} {\bibinfo {author} {\bibfnamefont {S.}~\bibnamefont
  {Emori}}, \bibinfo {author} {\bibfnamefont {U.}~\bibnamefont {Bauer}},
  \bibinfo {author} {\bibfnamefont {S.-M.}\ \bibnamefont {Ahn}}, \bibinfo
  {author} {\bibfnamefont {E.}~\bibnamefont {Martinez}}, \ and\ \bibinfo
  {author} {\bibfnamefont {G.~S.~D.}\ \bibnamefont {Beach}},\ }\href@noop {}
  {\bibfield  {journal} {\bibinfo  {journal} {Nat. Mater.},\ }\textbf {\bibinfo
  {volume} {12}},\ \bibinfo {pages} {611} (\bibinfo {year} {2013})}\BibitemShut
  {NoStop}%
\bibitem [{\citenamefont {Ryu}\ \emph {et~al.}(2013)\citenamefont {Ryu},
  \citenamefont {Thomas}, \citenamefont {Yang},\ and\ \citenamefont
  {Parkin}}]{Ryu2013}%
  \BibitemOpen
  \bibfield  {author} {\bibinfo {author} {\bibfnamefont {K.-S.}\ \bibnamefont
  {Ryu}}, \bibinfo {author} {\bibfnamefont {L.}~\bibnamefont {Thomas}},
  \bibinfo {author} {\bibfnamefont {S.-H.}\ \bibnamefont {Yang}}, \ and\
  \bibinfo {author} {\bibfnamefont {S.}~\bibnamefont {Parkin}},\ }\href@noop {}
  {\bibfield  {journal} {\bibinfo  {journal} {Nat. Nano.},\ }\textbf {\bibinfo
  {volume} {8}},\ \bibinfo {pages} {527} (\bibinfo {year} {2013})}\BibitemShut
  {NoStop}%
\bibitem [{\citenamefont {Martinez}\ \emph {et~al.}(2013)\citenamefont
  {Martinez}, \citenamefont {Emori},\ and\ \citenamefont
  {Beach}}]{martinez2013}%
  \BibitemOpen
  \bibfield  {author} {\bibinfo {author} {\bibfnamefont {E.}~\bibnamefont
  {Martinez}}, \bibinfo {author} {\bibfnamefont {S.}~\bibnamefont {Emori}}, \
  and\ \bibinfo {author} {\bibfnamefont {G.~S.~D.}\ \bibnamefont {Beach}},\
  }\Doi {http://dx.doi.org/10.1063/1.4818723} {\bibfield  {journal} {\bibinfo
  {journal} {Applied Physics Letters},\ }\textbf {\bibinfo {volume} {103}},\
  \bibinfo {eid} {072406} (\bibinfo {year} {2013})}\BibitemShut {NoStop}%
\bibitem [{\citenamefont {Torrejon}\ \emph {et~al.}(2014)\citenamefont
  {Torrejon}, \citenamefont {Kim}, \citenamefont {Sinha}, \citenamefont
  {Mitani}, \citenamefont {Hayashi}, \citenamefont {Yamanouchi},\ and\
  \citenamefont {Ohno}}]{Torrejon2014}%
  \BibitemOpen
  \bibfield  {author} {\bibinfo {author} {\bibfnamefont {J.}~\bibnamefont
  {Torrejon}}, \bibinfo {author} {\bibfnamefont {J.}~\bibnamefont {Kim}},
  \bibinfo {author} {\bibfnamefont {J.}~\bibnamefont {Sinha}}, \bibinfo
  {author} {\bibfnamefont {S.}~\bibnamefont {Mitani}}, \bibinfo {author}
  {\bibfnamefont {M.}~\bibnamefont {Hayashi}}, \bibinfo {author} {\bibfnamefont
  {M.}~\bibnamefont {Yamanouchi}}, \ and\ \bibinfo {author} {\bibfnamefont
  {H.}~\bibnamefont {Ohno}},\ }\href {http://dx.doi.org/10.1038/ncomms5655}
  {\bibfield  {journal} {\bibinfo  {journal} {Nat. Commun.},\ }\textbf
  {\bibinfo {volume} {5}} (\bibinfo {year} {2014})}\BibitemShut {NoStop}%
\bibitem [{\citenamefont {Garello}\ \emph {et~al.}(2013)\citenamefont
  {Garello}, \citenamefont {Miron}, \citenamefont {Avci}, \citenamefont
  {Freimuth}, \citenamefont {Mokrousov}, \citenamefont {Bl$\ddot{\rm u}$gel},
  \citenamefont {Auffret}, \citenamefont {Boulle}, \citenamefont {Gaudin},\
  and\ \citenamefont {Gambardella}}]{Garello2013}%
  \BibitemOpen
  \bibfield  {author} {\bibinfo {author} {\bibfnamefont {K.}~\bibnamefont
  {Garello}}, \bibinfo {author} {\bibfnamefont {I.~M.}\ \bibnamefont {Miron}},
  \bibinfo {author} {\bibfnamefont {C.~O.}\ \bibnamefont {Avci}}, \bibinfo
  {author} {\bibfnamefont {F.}~\bibnamefont {Freimuth}}, \bibinfo {author}
  {\bibfnamefont {Y.}~\bibnamefont {Mokrousov}}, \bibinfo {author}
  {\bibfnamefont {S.}~\bibnamefont {Bl$\ddot{\rm u}$gel}}, \bibinfo {author}
  {\bibfnamefont {S.}~\bibnamefont {Auffret}}, \bibinfo {author} {\bibfnamefont
  {O.}~\bibnamefont {Boulle}}, \bibinfo {author} {\bibfnamefont
  {G.}~\bibnamefont {Gaudin}}, \ and\ \bibinfo {author} {\bibfnamefont
  {P.}~\bibnamefont {Gambardella}},\ }\href@noop {} {\bibfield  {journal}
  {\bibinfo  {journal} {Nat. Nano.},\ }\textbf {\bibinfo {volume} {8}},\
  \bibinfo {pages} {587} (\bibinfo {year} {2013})}\BibitemShut {NoStop}%
\bibitem [{\citenamefont {Va\v{n}atka}\ \emph {et~al.}(2015)\citenamefont
  {Va\v{n}atka}, \citenamefont {Rojas-S\'{a}nchez}, \citenamefont {Vogel},
  \citenamefont {Bonfim}, \citenamefont {Belmeguenai}, \citenamefont
  {Roussign\'{e}}, \citenamefont {Stashkevich}, \citenamefont {Thiaville},\
  and\ \citenamefont {Pizzini}}]{Pizzini2015}%
  \BibitemOpen
  \bibfield  {author} {\bibinfo {author} {\bibfnamefont {M.}~\bibnamefont
  {Va\v{n}atka}}, \bibinfo {author} {\bibfnamefont {J.-C.}\ \bibnamefont
  {Rojas-S\'{a}nchez}}, \bibinfo {author} {\bibfnamefont {J.}~\bibnamefont
  {Vogel}}, \bibinfo {author} {\bibfnamefont {M.}~\bibnamefont {Bonfim}},
  \bibinfo {author} {\bibfnamefont {M.}~\bibnamefont {Belmeguenai}}, \bibinfo
  {author} {\bibfnamefont {Y.}~\bibnamefont {Roussign\'{e}}}, \bibinfo {author}
  {\bibfnamefont {A.}~\bibnamefont {Stashkevich}}, \bibinfo {author}
  {\bibfnamefont {A.}~\bibnamefont {Thiaville}}, \ and\ \bibinfo {author}
  {\bibfnamefont {S.}~\bibnamefont {Pizzini}},\ }\href
  {http://stacks.iop.org/0953-8984/27/i=32/a=326002} {\bibfield  {journal}
  {\bibinfo  {journal} {J. Phys.: Condensed Matter},\ }\textbf {\bibinfo
  {volume} {27}},\ \bibinfo {pages} {326002} (\bibinfo {year}
  {2015})}\BibitemShut {NoStop}%
\bibitem [{\citenamefont {Lavrijsen}\ \emph {et~al.}(2015)\citenamefont
  {Lavrijsen}, \citenamefont {Hartmann}, \citenamefont {van~den Brink},
  \citenamefont {Yin}, \citenamefont {Barcones}, \citenamefont {Duine},
  \citenamefont {Verheijen}, \citenamefont {Swagten},\ and\ \citenamefont
  {Koopmans}}]{Lavrijsen2015}%
  \BibitemOpen
  \bibfield  {author} {\bibinfo {author} {\bibfnamefont {R.}~\bibnamefont
  {Lavrijsen}}, \bibinfo {author} {\bibfnamefont {D.~M.~F.}\ \bibnamefont
  {Hartmann}}, \bibinfo {author} {\bibfnamefont {A.}~\bibnamefont {van~den
  Brink}}, \bibinfo {author} {\bibfnamefont {Y.}~\bibnamefont {Yin}}, \bibinfo
  {author} {\bibfnamefont {B.}~\bibnamefont {Barcones}}, \bibinfo {author}
  {\bibfnamefont {R.~A.}\ \bibnamefont {Duine}}, \bibinfo {author}
  {\bibfnamefont {M.~A.}\ \bibnamefont {Verheijen}}, \bibinfo {author}
  {\bibfnamefont {H.~J.~M.}\ \bibnamefont {Swagten}}, \ and\ \bibinfo {author}
  {\bibfnamefont {B.}~\bibnamefont {Koopmans}},\ }\href@noop {} {\bibfield
  {journal} {\bibinfo  {journal} {Phys. Rev. B},\ }\textbf {\bibinfo {volume}
  {91}},\ \bibinfo {pages} {104414} (\bibinfo {year} {2015})}\BibitemShut
  {NoStop}%
\bibitem [{\citenamefont {Di}\ \emph {et~al.}(2015)\citenamefont {Di},
  \citenamefont {Zhang}, \citenamefont {Lim}, \citenamefont {Ng}, \citenamefont
  {Kuok}, \citenamefont {Yu}, \citenamefont {Yoon}, \citenamefont {Qiu},\ and\
  \citenamefont {Yang}}]{KaiPRL2015}%
  \BibitemOpen
  \bibfield  {author} {\bibinfo {author} {\bibfnamefont {K.}~\bibnamefont
  {Di}}, \bibinfo {author} {\bibfnamefont {V.~L.}\ \bibnamefont {Zhang}},
  \bibinfo {author} {\bibfnamefont {H.~S.}\ \bibnamefont {Lim}}, \bibinfo
  {author} {\bibfnamefont {S.~C.}\ \bibnamefont {Ng}}, \bibinfo {author}
  {\bibfnamefont {M.~H.}\ \bibnamefont {Kuok}}, \bibinfo {author}
  {\bibfnamefont {J.}~\bibnamefont {Yu}}, \bibinfo {author} {\bibfnamefont
  {J.}~\bibnamefont {Yoon}}, \bibinfo {author} {\bibfnamefont {X.}~\bibnamefont
  {Qiu}}, \ and\ \bibinfo {author} {\bibfnamefont {H.}~\bibnamefont {Yang}},\
  }\href@noop {} {\bibfield  {journal} {\bibinfo  {journal} {Phys. Rev.
  Lett.},\ }\textbf {\bibinfo {volume} {114}},\ \bibinfo {pages} {047201}
  (\bibinfo {year} {2015})}\BibitemShut {NoStop}%
\bibitem [{\citenamefont {Nembach}\ \emph {et~al.}(2015)\citenamefont
  {Nembach}, \citenamefont {Shaw}, \citenamefont {Weiler}, \citenamefont
  {Ju\'e},\ and\ \citenamefont {Silva}}]{Nembach2015}%
  \BibitemOpen
  \bibfield  {author} {\bibinfo {author} {\bibfnamefont {H.~T.}\ \bibnamefont
  {Nembach}}, \bibinfo {author} {\bibfnamefont {J.~M.}\ \bibnamefont {Shaw}},
  \bibinfo {author} {\bibfnamefont {M.}~\bibnamefont {Weiler}}, \bibinfo
  {author} {\bibfnamefont {E.}~\bibnamefont {Ju\'e}}, \ and\ \bibinfo {author}
  {\bibfnamefont {T.~J.}\ \bibnamefont {Silva}},\ }\href@noop {} {\bibfield
  {journal} {\bibinfo  {journal} {Nat. Phys.},\ }\textbf {\bibinfo {volume}
  {11}},\ \bibinfo {pages} {825} (\bibinfo {year} {2015})}\BibitemShut
  {NoStop}%
\bibitem [{\citenamefont {Stashkevich}\ \emph {et~al.}(2015)\citenamefont
  {Stashkevich}, \citenamefont {Belmeguenai}, \citenamefont {Roussign\'e},
  \citenamefont {Cherif}, \citenamefont {Kostylev}, \citenamefont {Gabor},
  \citenamefont {Lacour}, \citenamefont {Tiusan},\ and\ \citenamefont
  {Hehn}}]{Stashkevich2015}%
  \BibitemOpen
  \bibfield  {author} {\bibinfo {author} {\bibfnamefont {A.~A.}\ \bibnamefont
  {Stashkevich}}, \bibinfo {author} {\bibfnamefont {M.}~\bibnamefont
  {Belmeguenai}}, \bibinfo {author} {\bibfnamefont {Y.}~\bibnamefont
  {Roussign\'e}}, \bibinfo {author} {\bibfnamefont {S.~M.}\ \bibnamefont
  {Cherif}}, \bibinfo {author} {\bibfnamefont {M.}~\bibnamefont {Kostylev}},
  \bibinfo {author} {\bibfnamefont {M.}~\bibnamefont {Gabor}}, \bibinfo
  {author} {\bibfnamefont {D.}~\bibnamefont {Lacour}}, \bibinfo {author}
  {\bibfnamefont {C.}~\bibnamefont {Tiusan}}, \ and\ \bibinfo {author}
  {\bibfnamefont {M.}~\bibnamefont {Hehn}},\ }\href@noop {} {\bibfield
  {journal} {\bibinfo  {journal} {Phys. Rev. B},\ }\textbf {\bibinfo {volume}
  {91}},\ \bibinfo {pages} {214409} (\bibinfo {year} {2015})}\BibitemShut
  {NoStop}%
\bibitem [{\citenamefont {Belmeguenai}\ \emph {et~al.}(2015)\citenamefont
  {Belmeguenai}, \citenamefont {Adam}, \citenamefont {Roussign\'e},
  \citenamefont {Eimer}, \citenamefont {Devolder}, \citenamefont {Kim},
  \citenamefont {Cherif}, \citenamefont {Stashkevich},\ and\ \citenamefont
  {Thiaville}}]{Belmeguenai2015}%
  \BibitemOpen
  \bibfield  {author} {\bibinfo {author} {\bibfnamefont {M.}~\bibnamefont
  {Belmeguenai}}, \bibinfo {author} {\bibfnamefont {J.-P.}\ \bibnamefont
  {Adam}}, \bibinfo {author} {\bibfnamefont {Y.}~\bibnamefont {Roussign\'e}},
  \bibinfo {author} {\bibfnamefont {S.}~\bibnamefont {Eimer}}, \bibinfo
  {author} {\bibfnamefont {T.}~\bibnamefont {Devolder}}, \bibinfo {author}
  {\bibfnamefont {J.-V.}\ \bibnamefont {Kim}}, \bibinfo {author} {\bibfnamefont
  {S.~M.}\ \bibnamefont {Cherif}}, \bibinfo {author} {\bibfnamefont
  {A.}~\bibnamefont {Stashkevich}}, \ and\ \bibinfo {author} {\bibfnamefont
  {A.}~\bibnamefont {Thiaville}},\ }\href@noop {} {\bibfield  {journal}
  {\bibinfo  {journal} {Phys. Rev. B},\ }\textbf {\bibinfo {volume} {91}},\
  \bibinfo {pages} {180405} (\bibinfo {year} {2015})}\BibitemShut {NoStop}%
\bibitem [{\citenamefont {Tetienne}\ \emph {et~al.}(2015)\citenamefont
  {Tetienne}, \citenamefont {Hingant}, \citenamefont {Mart{\'\i}nez},
  \citenamefont {Rohart}, \citenamefont {Thiaville}, \citenamefont
  {Herrera~Diez}, \citenamefont {Garcia}, \citenamefont {Adam}, \citenamefont
  {Kim}, \citenamefont {Roch}, \citenamefont {Miron}, \citenamefont {Gaudin},
  \citenamefont {Vila}, \citenamefont {Ocker}, \citenamefont {Ravelosona},\
  and\ \citenamefont {Jacques}}]{Tetienne2015}%
  \BibitemOpen
  \bibfield  {author} {\bibinfo {author} {\bibfnamefont {J.~P.}\ \bibnamefont
  {Tetienne}}, \bibinfo {author} {\bibfnamefont {T.}~\bibnamefont {Hingant}},
  \bibinfo {author} {\bibfnamefont {L.~J.}\ \bibnamefont {Mart{\'\i}nez}},
  \bibinfo {author} {\bibfnamefont {S.}~\bibnamefont {Rohart}}, \bibinfo
  {author} {\bibfnamefont {A.}~\bibnamefont {Thiaville}}, \bibinfo {author}
  {\bibfnamefont {L.}~\bibnamefont {Herrera~Diez}}, \bibinfo {author}
  {\bibfnamefont {K.}~\bibnamefont {Garcia}}, \bibinfo {author} {\bibfnamefont
  {J.~P.}\ \bibnamefont {Adam}}, \bibinfo {author} {\bibfnamefont {J.~V.}\
  \bibnamefont {Kim}}, \bibinfo {author} {\bibfnamefont {J.~F.}\ \bibnamefont
  {Roch}}, \bibinfo {author} {\bibfnamefont {I.~M.}\ \bibnamefont {Miron}},
  \bibinfo {author} {\bibfnamefont {G.}~\bibnamefont {Gaudin}}, \bibinfo
  {author} {\bibfnamefont {L.}~\bibnamefont {Vila}}, \bibinfo {author}
  {\bibfnamefont {B.}~\bibnamefont {Ocker}}, \bibinfo {author} {\bibfnamefont
  {D.}~\bibnamefont {Ravelosona}}, \ and\ \bibinfo {author} {\bibfnamefont
  {V.}~\bibnamefont {Jacques}},\ }\href {http://dx.doi.org/10.1038/ncomms7733}
  {\bibfield  {journal} {\bibinfo  {journal} {Nat. Commun.},\ }\textbf
  {\bibinfo {volume} {6}} (\bibinfo {year} {2015})}\BibitemShut {NoStop}%
\bibitem [{\citenamefont {Tetienne}\ \emph
  {et~al.}(2014){\natexlab{a}}\citenamefont {Tetienne}, \citenamefont
  {Hingant}, \citenamefont {Rondin}, \citenamefont {Rohart}, \citenamefont
  {Thiaville}, \citenamefont {Ju\'e}, \citenamefont {Gaudin}, \citenamefont
  {Roch},\ and\ \citenamefont {Jacques}}]{TetienneJAP2014}%
  \BibitemOpen
  \bibfield  {author} {\bibinfo {author} {\bibfnamefont {J.-P.}\ \bibnamefont
  {Tetienne}}, \bibinfo {author} {\bibfnamefont {T.}~\bibnamefont {Hingant}},
  \bibinfo {author} {\bibfnamefont {L.}~\bibnamefont {Rondin}}, \bibinfo
  {author} {\bibfnamefont {S.}~\bibnamefont {Rohart}}, \bibinfo {author}
  {\bibfnamefont {A.}~\bibnamefont {Thiaville}}, \bibinfo {author}
  {\bibfnamefont {E.}~\bibnamefont {Ju\'e}}, \bibinfo {author} {\bibfnamefont
  {G.}~\bibnamefont {Gaudin}}, \bibinfo {author} {\bibfnamefont {J.-F.}\
  \bibnamefont {Roch}}, \ and\ \bibinfo {author} {\bibfnamefont
  {V.}~\bibnamefont {Jacques}},\ }\href@noop {} {\bibfield  {journal} {\bibinfo
   {journal} {J. Appl. Phys.},\ }\textbf {\bibinfo {volume} {115}},\ \bibinfo
  {pages} {17D501} (\bibinfo {year} {2014}{\natexlab{a}})}\BibitemShut
  {NoStop}%
\bibitem [{\citenamefont {Rondin}\ \emph {et~al.}(2012)\citenamefont {Rondin},
  \citenamefont {Tetienne}, \citenamefont {Spinicelli}, \citenamefont
  {Dal~Savio}, \citenamefont {Karrai}, \citenamefont {Dantelle}, \citenamefont
  {Thiaville}, \citenamefont {Rohart}, \citenamefont {Roch},\ and\
  \citenamefont {Jacques}}]{Rondin2012}%
  \BibitemOpen
  \bibfield  {author} {\bibinfo {author} {\bibfnamefont {L.}~\bibnamefont
  {Rondin}}, \bibinfo {author} {\bibfnamefont {J.-P.}\ \bibnamefont
  {Tetienne}}, \bibinfo {author} {\bibfnamefont {P.}~\bibnamefont
  {Spinicelli}}, \bibinfo {author} {\bibfnamefont {C.}~\bibnamefont
  {Dal~Savio}}, \bibinfo {author} {\bibfnamefont {K.}~\bibnamefont {Karrai}},
  \bibinfo {author} {\bibfnamefont {G.}~\bibnamefont {Dantelle}}, \bibinfo
  {author} {\bibfnamefont {A.}~\bibnamefont {Thiaville}}, \bibinfo {author}
  {\bibfnamefont {S.}~\bibnamefont {Rohart}}, \bibinfo {author} {\bibfnamefont
  {J.-F.}\ \bibnamefont {Roch}}, \ and\ \bibinfo {author} {\bibfnamefont
  {V.}~\bibnamefont {Jacques}},\ }\href@noop {} {\bibfield  {journal} {\bibinfo
   {journal} {Appl. Phys. Lett.},\ }\textbf {\bibinfo {volume} {100}},\
  \bibinfo {eid} {153118} (\bibinfo {year} {2012})}\BibitemShut {NoStop}%
\bibitem [{\citenamefont {Rondin}\ \emph {et~al.}(2014)\citenamefont {Rondin},
  \citenamefont {Tetienne}, \citenamefont {Hingant}, \citenamefont {Roch},
  \citenamefont {Maletinsky},\ and\ \citenamefont {Jacques}}]{Rondin2014}%
  \BibitemOpen
  \bibfield  {author} {\bibinfo {author} {\bibfnamefont {L.}~\bibnamefont
  {Rondin}}, \bibinfo {author} {\bibfnamefont {J.-P.}\ \bibnamefont
  {Tetienne}}, \bibinfo {author} {\bibfnamefont {T.}~\bibnamefont {Hingant}},
  \bibinfo {author} {\bibfnamefont {J.-F.}\ \bibnamefont {Roch}}, \bibinfo
  {author} {\bibfnamefont {P.}~\bibnamefont {Maletinsky}}, \ and\ \bibinfo
  {author} {\bibfnamefont {V.}~\bibnamefont {Jacques}},\ }\href
  {http://stacks.iop.org/0034-4885/77/i=5/a=056503} {\bibfield  {journal}
  {\bibinfo  {journal} {Rep. Prog. Phys.},\ }\textbf {\bibinfo {volume} {77}},\
  \bibinfo {pages} {056503} (\bibinfo {year} {2014})}\BibitemShut {NoStop}%
\bibitem [{\citenamefont {Hingant}\ \emph {et~al.}(2015)\citenamefont
  {Hingant}, \citenamefont {Tetienne}, \citenamefont {Mart\'{\i}nez},
  \citenamefont {Garcia}, \citenamefont {Ravelosona}, \citenamefont {Roch},\
  and\ \citenamefont {Jacques}}]{Hingant2015}%
  \BibitemOpen
  \bibfield  {author} {\bibinfo {author} {\bibfnamefont {T.}~\bibnamefont
  {Hingant}}, \bibinfo {author} {\bibfnamefont {J.-P.}\ \bibnamefont
  {Tetienne}}, \bibinfo {author} {\bibfnamefont {L.~J.}\ \bibnamefont
  {Mart\'{\i}nez}}, \bibinfo {author} {\bibfnamefont {K.}~\bibnamefont
  {Garcia}}, \bibinfo {author} {\bibfnamefont {D.}~\bibnamefont {Ravelosona}},
  \bibinfo {author} {\bibfnamefont {J.-F.}\ \bibnamefont {Roch}}, \ and\
  \bibinfo {author} {\bibfnamefont {V.}~\bibnamefont {Jacques}},\ }\Doi
  {10.1103/PhysRevApplied.4.014003} {\bibfield  {journal} {\bibinfo  {journal}
  {Phys. Rev. Applied},\ }\textbf {\bibinfo {volume} {4}},\ \bibinfo {pages}
  {014003} (\bibinfo {year} {2015})}\BibitemShut {NoStop}%
\bibitem [{\citenamefont {Yamanouchi}\ \emph {et~al.}(2011)\citenamefont
  {Yamanouchi}, \citenamefont {Jander}, \citenamefont {Dhagat}, \citenamefont
  {Ikeda}, \citenamefont {Matsukura},\ and\ \citenamefont
  {Ohno}}]{Yamanouchi2011}%
  \BibitemOpen
  \bibfield  {author} {\bibinfo {author} {\bibfnamefont {M.}~\bibnamefont
  {Yamanouchi}}, \bibinfo {author} {\bibfnamefont {A.}~\bibnamefont {Jander}},
  \bibinfo {author} {\bibfnamefont {P.}~\bibnamefont {Dhagat}}, \bibinfo
  {author} {\bibfnamefont {S.}~\bibnamefont {Ikeda}}, \bibinfo {author}
  {\bibfnamefont {F.}~\bibnamefont {Matsukura}}, \ and\ \bibinfo {author}
  {\bibfnamefont {H.}~\bibnamefont {Ohno}},\ }\href@noop {} {\bibfield
  {journal} {\bibinfo  {journal} {IEEE Magn. Lett.},\ }\textbf {\bibinfo
  {volume} {2}},\ \bibinfo {pages} {3000304} (\bibinfo {year}
  {2011})}\BibitemShut {NoStop}%
\bibitem [{\citenamefont {Sinha}\ \emph {et~al.}(2013)\citenamefont {Sinha},
  \citenamefont {Hayashi}, \citenamefont {Kellock}, \citenamefont {Fukami},
  \citenamefont {Yamanouchi}, \citenamefont {Sato}, \citenamefont {Ikeda},
  \citenamefont {Mitani}, \citenamefont {Yang}, \citenamefont {Parkin},\ and\
  \citenamefont {Ohno}}]{Sinha2013}%
  \BibitemOpen
  \bibfield  {author} {\bibinfo {author} {\bibfnamefont {J.}~\bibnamefont
  {Sinha}}, \bibinfo {author} {\bibfnamefont {M.}~\bibnamefont {Hayashi}},
  \bibinfo {author} {\bibfnamefont {A.~J.}\ \bibnamefont {Kellock}}, \bibinfo
  {author} {\bibfnamefont {S.}~\bibnamefont {Fukami}}, \bibinfo {author}
  {\bibfnamefont {M.}~\bibnamefont {Yamanouchi}}, \bibinfo {author}
  {\bibfnamefont {H.}~\bibnamefont {Sato}}, \bibinfo {author} {\bibfnamefont
  {S.}~\bibnamefont {Ikeda}}, \bibinfo {author} {\bibfnamefont
  {S.}~\bibnamefont {Mitani}}, \bibinfo {author} {\bibfnamefont
  {S.}~\bibnamefont {Yang}}, \bibinfo {author} {\bibfnamefont {S.~S.~P.}\
  \bibnamefont {Parkin}}, \ and\ \bibinfo {author} {\bibfnamefont
  {H.}~\bibnamefont {Ohno}},\ }\href@noop {} {\bibfield  {journal} {\bibinfo
  {journal} {Appl. Phys. Lett.},\ }\textbf {\bibinfo {volume} {102}},\ \bibinfo
  {pages} {242405} (\bibinfo {year} {2013})}\BibitemShut {NoStop}%
\bibitem [{\citenamefont {Soucaille}\ \emph {et~al.}(2016)\citenamefont
  {Soucaille}, \citenamefont {Belmeguenai}, \citenamefont {Torrejon},
  \citenamefont {Kim}, \citenamefont {Devolder}, \citenamefont {Roussign\'e},
  \citenamefont {Ch\'erif}, \citenamefont {Stashkevich}, \citenamefont
  {Hayashi},\ and\ \citenamefont {Adam}}]{Soucaille2016}%
  \BibitemOpen
  \bibfield  {author} {\bibinfo {author} {\bibfnamefont {R.}~\bibnamefont
  {Soucaille}}, \bibinfo {author} {\bibfnamefont {M.}~\bibnamefont
  {Belmeguenai}}, \bibinfo {author} {\bibfnamefont {J.}~\bibnamefont
  {Torrejon}}, \bibinfo {author} {\bibfnamefont {J.-V.}\ \bibnamefont {Kim}},
  \bibinfo {author} {\bibfnamefont {T.}~\bibnamefont {Devolder}}, \bibinfo
  {author} {\bibfnamefont {Y.}~\bibnamefont {Roussign\'e}}, \bibinfo {author}
  {\bibfnamefont {S.-M.}\ \bibnamefont {Ch\'erif}}, \bibinfo {author}
  {\bibfnamefont {A.~A.}\ \bibnamefont {Stashkevich}}, \bibinfo {author}
  {\bibfnamefont {M.}~\bibnamefont {Hayashi}}, \ and\ \bibinfo {author}
  {\bibfnamefont {J.-P.}\ \bibnamefont {Adam}},\ }\href@noop {} {\bibfield
  {journal} {\bibinfo  {journal} {Preprint arXiv:1604.05475}} (\bibinfo {year}
  {2016})}\BibitemShut {NoStop}%
\bibitem [{\citenamefont {Burrowes}\ \emph {et~al.}(2013)\citenamefont
  {Burrowes}, \citenamefont {Vernier}, \citenamefont {Adam}, \citenamefont
  {Herrera~Diez}, \citenamefont {Garcia}, \citenamefont {Barisic},
  \citenamefont {Agnus}, \citenamefont {Eimer}, \citenamefont {Kim},
  \citenamefont {Devolder}, \citenamefont {Lamperti}, \citenamefont {Mantovan},
  \citenamefont {Ockert}, \citenamefont {Fullerton},\ and\ \citenamefont
  {Ravelosona}}]{Burrowes}%
  \BibitemOpen
  \bibfield  {author} {\bibinfo {author} {\bibfnamefont {C.}~\bibnamefont
  {Burrowes}}, \bibinfo {author} {\bibfnamefont {N.}~\bibnamefont {Vernier}},
  \bibinfo {author} {\bibfnamefont {J.-P.}\ \bibnamefont {Adam}}, \bibinfo
  {author} {\bibfnamefont {L.}~\bibnamefont {Herrera~Diez}}, \bibinfo {author}
  {\bibfnamefont {K.}~\bibnamefont {Garcia}}, \bibinfo {author} {\bibfnamefont
  {I.}~\bibnamefont {Barisic}}, \bibinfo {author} {\bibfnamefont
  {G.}~\bibnamefont {Agnus}}, \bibinfo {author} {\bibfnamefont
  {S.}~\bibnamefont {Eimer}}, \bibinfo {author} {\bibfnamefont {J.-V.}\
  \bibnamefont {Kim}}, \bibinfo {author} {\bibfnamefont {T.}~\bibnamefont
  {Devolder}}, \bibinfo {author} {\bibfnamefont {A.}~\bibnamefont {Lamperti}},
  \bibinfo {author} {\bibfnamefont {R.}~\bibnamefont {Mantovan}}, \bibinfo
  {author} {\bibfnamefont {B.}~\bibnamefont {Ockert}}, \bibinfo {author}
  {\bibfnamefont {E.~E.}\ \bibnamefont {Fullerton}}, \ and\ \bibinfo {author}
  {\bibfnamefont {D.}~\bibnamefont {Ravelosona}},\ }\href@noop {} {\bibfield
  {journal} {\bibinfo  {journal} {Appl. Phys. Lett.},\ }\textbf {\bibinfo
  {volume} {103}},\ \bibinfo {eid} {182401} (\bibinfo {year}
  {2013})}\BibitemShut {NoStop}%
\bibitem [{\citenamefont {Yamada}\ \emph {et~al.}(2011)\citenamefont {Yamada},
  \citenamefont {Jamet}, \citenamefont {Nakatani}, \citenamefont {Mougin},
  \citenamefont {Thiaville}, \citenamefont {Ono},\ and\ \citenamefont
  {Ferr\'e}}]{Yamada}%
  \BibitemOpen
  \bibfield  {author} {\bibinfo {author} {\bibfnamefont {K.}~\bibnamefont
  {Yamada}}, \bibinfo {author} {\bibfnamefont {J.-P.}\ \bibnamefont {Jamet}},
  \bibinfo {author} {\bibfnamefont {Y.}~\bibnamefont {Nakatani}}, \bibinfo
  {author} {\bibfnamefont {A.}~\bibnamefont {Mougin}}, \bibinfo {author}
  {\bibfnamefont {A.}~\bibnamefont {Thiaville}}, \bibinfo {author}
  {\bibfnamefont {T.}~\bibnamefont {Ono}}, \ and\ \bibinfo {author}
  {\bibfnamefont {J.}~\bibnamefont {Ferr\'e}},\ }\href@noop {} {\bibfield
  {journal} {\bibinfo  {journal} {Appl. Phys. Express},\ }\textbf {\bibinfo
  {volume} {4}},\ \bibinfo {pages} {113001} (\bibinfo {year}
  {2011})}\BibitemShut {NoStop}%
\bibitem [{\citenamefont {Tetienne}\ \emph
  {et~al.}(2014){\natexlab{b}}\citenamefont {Tetienne}, \citenamefont
  {Hingant}, \citenamefont {Kim}, \citenamefont {Herrera~Diez}, \citenamefont
  {Adam}, \citenamefont {Garcia}, \citenamefont {Roch}, \citenamefont {Rohart},
  \citenamefont {Thiaville}, \citenamefont {Ravelosona},\ and\ \citenamefont
  {Jacques}}]{Tetienne2014}%
  \BibitemOpen
  \bibfield  {author} {\bibinfo {author} {\bibfnamefont {J.-P.}\ \bibnamefont
  {Tetienne}}, \bibinfo {author} {\bibfnamefont {T.}~\bibnamefont {Hingant}},
  \bibinfo {author} {\bibfnamefont {J.-V.}\ \bibnamefont {Kim}}, \bibinfo
  {author} {\bibfnamefont {L.}~\bibnamefont {Herrera~Diez}}, \bibinfo {author}
  {\bibfnamefont {J.-P.}\ \bibnamefont {Adam}}, \bibinfo {author}
  {\bibfnamefont {K.}~\bibnamefont {Garcia}}, \bibinfo {author} {\bibfnamefont
  {J.-F.}\ \bibnamefont {Roch}}, \bibinfo {author} {\bibfnamefont
  {S.}~\bibnamefont {Rohart}}, \bibinfo {author} {\bibfnamefont
  {A.}~\bibnamefont {Thiaville}}, \bibinfo {author} {\bibfnamefont
  {D.}~\bibnamefont {Ravelosona}}, \ and\ \bibinfo {author} {\bibfnamefont
  {V.}~\bibnamefont {Jacques}},\ }\href@noop {} {\bibfield  {journal} {\bibinfo
   {journal} {Science},\ }\textbf {\bibinfo {volume} {344}},\ \bibinfo {pages}
  {1366} (\bibinfo {year} {2014}{\natexlab{b}})}\BibitemShut {NoStop}%
\end{thebibliography}%

\end{document}